\documentstyle[prc,aps,amssymb,epsf]{revtex}
\def\la{\langle}
\def\ra{\rangle}
\def\beq{\begin{equation}}
\def\eeq{\end{equation}}
\def\be{\begin{eqnarray}}
\def\ee{\end{eqnarray}}
\def\hs{\hat{s}}
\def\htm{\hat{t}}
\def\hu{\hat{u}}

\def\kav{\la k_T\ra}
\def\k2av{\la k_T^2\ra}
\newcommand{\f}[2]{\frac{#1}{#2}}
\newcommand{\dd}{ {\textrm d}}
\draft
 
\begin{document}

\title{High-$p_T$ pion and kaon production in relativistic nuclear collisions} 
\author{Yi Zhang\thanks{electronic mail: yzhang1@cnr4.physics.kent.edu}
 and George Fai\thanks{electronic mail: fai@cnrred.kent.edu}}
\address{Center for Nuclear Research, Department of Physics\\
         Kent State University, Kent, OH 44242}
\author{G{\'a}bor Papp\thanks{electronic mail: pg@ludens.elte.hu}}
\address{HAS Research Group for Theoretical Physics\\
E{\"o}tv{\"o}s University, P{\'a}zm{\'a}ny P. 1/A, Budapest 1117, Hungary}
\author{Gergely G. Barnaf{\"o}ldi\thanks{electronic mail: bgergely@rmki.kfki.hu}
 and P{\'e}ter L{\'e}vai\thanks{electronic mail: plevai@rmki.kfki.hu}}
\address{KFKI Research Institute for Particle and Nuclear Physics,
P.O. Box 49, Budapest 1525, Hungary}

\date{\today}

\maketitle

\vspace*{-6.5cm}
\begin{flushright} 
{KSUCNR-102-01}

\end{flushright}
\vspace*{5.5cm}
 
\begin{abstract}
High-$p_T$ pion and kaon production is studied in 
relativistic proton-proton, proton-nucleus, and nucleus-nucleus 
collisions in a wide energy range. Cross sections
are calculated based on perturbative QCD, augmented by a 
phenomenological transverse momentum distribution of partons
(``intrinsic $k_T$''). 
An energy dependent width of the transverse momentum distribution 
is extracted from pion and charged hadron production data in 
proton-proton/proton-antiproton collisions. Effects of 
multiscattering and shadowing in the strongly interacting 
medium are taken into account. Enhancement of the transverse momentum 
width is introduced and parameterized to explain the Cronin effect. 
In collisions between heavy nuclei, the model over-predicts central 
pion production cross sections (more significantly at higher energies), 
hinting at the presence of jet quenching. Predictions are made
for proton-nucleus and nucleus-nucleus collisions at RHIC energies.    
\end{abstract}
 
%\begin{narrowtext}
%\newpage
\section{Introduction}
\label{sec_intro}
As the bombarding energy is increased, high--$p_T$ particle production becomes 
a prominent feature of nuclear collisions, as evidenced recently by the data 
beginning to emerge from the Relativistic Heavy Ion Collider (RHIC)\cite{QM01}.
These data call for concentrated theoretical efforts to overcome the challenges
presented by the high-energy and many-body features of nuclear collisions 
at center-of-mass energies of 100 AGeV $\lesssim \sqrt{s} \lesssim$ 200 AGeV.
Relying on asymptotic freedom, perturbative quantum 
chromodynamics (pQCD) can be applied at sufficiently high energies. 
To use pQCD as a practical tool,
one takes advantage of the factorization theorem, which provides a simple 
connection between the level of observed particles and that of the 
underlying quark-gluon (parton) structure. Briefly,
the observable cross sections are expressed in terms of a convolution of 
partonic cross sections with parton distribution functions (PDFs) 
and fragmentation functions (FFs), which encode some of the perturbatively 
non-calculable low-energy aspects of the physics. 
The resulting calculational scheme adopted here is referred to as the
pQCD-improved parton model~\cite{owens87,FF95}.
To assure the validity
of a perturbative treatment, we limit the discussion to {\it hard}
particle production, requiring the transverse momentum of the
inclusively measured produced particle
to be above a certain threshold, $p_T \geq p_{T0}$. Typical values of $p_{T0}$
are around 1 -- 3 GeV in the literature.  

Hard pion and kaon production in proton-proton ($pp$) and 
proton-nucleus ($pA$) collisions at these energies is in itself of interest,
but it is also imperative to study these reactions 
as a step in the process of understanding hard $\pi$ and $K$
production in nucleus-nucleus ($AA$) collisions. The necessity of this 
systematic approach has been made particularly clear by the high-visibility 
parallel developments in the subfield of $J/\psi$ production\cite{satz01,leit00}.
Significant amounts of experimental data on $\pi$ and $K$ production in $pp$ 
collisions 
are available in the energy range 20 GeV $\lesssim \sqrt{s} \lesssim$ 60~GeV
\cite{cronin75,antreasyan79,R806,CCRS,E605pi,E605pib,ISR}.   
Higher energies are less thoroughly explored, with mostly calorimetric data 
on total hadron production (in lieu of the identified pion and kaon spectra at
lower energies). In the present paper, we use 
CERN UA1\cite{UA1arn,UA1alb,UA1boc} and Tevatron CDF\cite{CDF}  
data from $p{\overline p} \rightarrow h^\pm$ reactions to bracket 
the RHIC energy range from above.

The presence of nuclear effects modifies the predictions of the  
pQCD-improved parton model for hard particle production in
$pA$ collisions, as indicated by available $pA$ hadron production 
data\cite{cronin75,antreasyan79,E605pib,brown96,ApaBep}. 
While hard particle production (usually disregarding nuclear 
effects) has been used as a testing ground for pQCD, the nuclear 
modifications are crucially important for RHIC experiments and planned
nuclear experiments at the Large Hadron Collider (LHC). In particular, 
to assess the effects of jet quenching\cite{plumer,baier98,gyulassy00}, 
precise background 
calculations and the knowledge of the gluon density of the medium responsible 
for the energy loss are necessary\cite{wang01;a}.

In a recent Rapid Communication\cite{plf00}, we suggested a specific form 
(``saturation'') of the nuclear modification for the cross section enhancement 
observed at high $p_T$ in $pA$ collisions over what would be expected based
on a simple scaling of the appropriate $pp$ cross section 
(Cronin effect)\cite{cronin75,antreasyan79}. 
The present paper provides a more detailed account of our work, incorporating 
not only neutral but 
also charged $\pi$ and $K$ production, with applications and predictions at 
RHIC energies and for $AA$ collisions. The saturation prescription \cite{plf00} 
is looked upon as a limiting case, with the opposite limit corresponding 
to the participation of all available nucleons in the enhancement of 
the transverse-momentum width (see Section \ref{sec_nu}). We study the 
dependence of the results on the possible choices between these limits, 
and suggest a connection of the preferred value of the number of
semi-hard (momentum transfer $\sim$ 0.5 GeV) 
collisions to the results of lower-energy 
experiments\cite{cole,Chem,Cole01,Veres01}. Further measurements
are needed to clarify the physical picture, and we argue 
that the planned $pA$ runs at RHIC\cite{pARHIC} are likely to play a very 
important role in this regard.  

A similar effort (with somewhat different emphasis)
was published recently\cite{wang01;b}. Initial RHIC 
data and the interest in jet quenching warrant an independent detailed 
study of the pQCD background to which jet quenching should be applied 
to achieve agreement with central $AA$ collision data. Our work differs 
from Ref \cite{wang01;b} on the basic pQCD level in the choice of scales and
in the use of the newly-published  KKP fragmentation functions\cite{KKP}.
Furthermore, we take the position of consistently using leading order (LO)
pQCD in the present paper, without introducing a so-called `K~factor', which we  
found to be energy and transverse momentum dependent in another 
publication\cite{bflpz01}. In this way we have fewer parameters, and
the burden of obtaining a good
fit to the $pp$ data rests entirely on the transverse momentum 
distribution, without mixing the effects 
of the K factor with those of 
the intrinsic transverse momentum. We determine the energy-dependent
width of the transverse momentum distribution by fitting $pp$ data,
in contrast to the prescribed scale dependence used in Ref. \cite{wang01;b}.

Our strategy in this paper is to first examine hard pion production
in $pp$ collisions up to $\sqrt{s} \lesssim 60$ GeV c.m. energy. We find that 
satisfactory agreement with the data can be achieved in this 
energy region in LO pQCD, utilizing
the width of the intrinsic transverse-momentum distribution of partons 
in the colliding nucleons. We treat this quantity (also measured in dijet
events, see e.g. Ref. \cite{E609}), as an energy-dependent 
nonperturbative parameter. This is discussed in Section 
\ref{sec_kT}. Having extracted the best value of the transverse-momentum
width from the $pp$ data, we apply these ideas to pion production in $pA$ 
collisions in the same energy range. Our choice for the effective number of 
semi-hard collisions is described in Section \ref{sec_nu}. We then discuss 
$AA$ collisions, and compare our results to CERN data on $\pi^0$ production 
in $S+S$ and $Pb+Pb$ collisions at $\sqrt{s} \sim 20$ AGeV. 
In Section \ref{sec_K} we move to $K^\pm$ production in $pp$, $pA$,
and $AA$ collisions at the same energies,
with the parameters as fixed above. A major step towards calculations 
at RHIC energies is covered in Section \ref{sec_RHICpp}, where we attempt to 
extend the energy range of the parameterization in terms of the width
of the transverse-momentum distribution of partons in the nucleon.
This is made difficult by the availability of  a smaller number of 
$p\bar{p}$ experiments, most of them at a significantly higher 
$\sqrt{s}$ than the RHIC energy domain, 
limited to the measurement of total charged hadron production. 
The implied uncertainties need to be kept in mind when we
display predictions for $pA$ and $AA$ collisions at RHIC energies,
in Sections \ref{sec_RHICpA} and \ref{sec_RHICAA}, respectively. 
Section \ref{sec_sum} contains our summary and conclusions. 
In the Appendix, we discuss the valence and sea contributions to 
quark fragmentation into  charged pions and kaons. We use $\hbar=c=1$ units.

%\newpage
\section{Pion production at center-of-mass energies below $\sim 60$ GeV}
\label{sec_pi}

In this Section, we first summarize the treatment of particle 
production in the pQCD-improved parton model,
listing various assumptions and ingredients. We introduce the 
intrinsic transverse-momentum distribution of partons and fit the 
width of this distribution to available $pp$ data in the 
energy range 20 GeV $\lesssim \sqrt{s} \lesssim$ 60 GeV in 
Subsection \ref{sec_kT}. A large body of
pion and kaon production data can be utilized in this energy interval
to extract information on the width of the transverse momentum distribution. 
In Subsection \ref{sec_nu} the Cronin effect is discussed in $pA$ collisions.
Subsection \ref{sec_AApi} deals with hard pion production in $AA$ collisions.

\subsection{Parton model and pQCD with intrinsic transverse momentum}
\label{sec_kT}     

The invariant cross section for the production of hadron $h$ in a $pp$ collision 
is described in the pQCD-improved parton model on the basis of the factorization 
theorem as a convolution\cite{FF95}: 
\beq
\label{hadX}
  E_{h}\f{\dd \sigma_h^{pp}}{\dd ^3p} =
        \sum_{abcd} \int\! \dd x_a \dd x_b \dd z_c \ f_{a/p}(x_a,Q^2)\
        f_{b/p}(x_b,Q^2)\ 
\f{\dd \sigma}{\dd \htm}(ab \to cd)\,
   \frac{D_{h/c}(z_c, Q'^2)}{\pi z_c^2} \, \hs \, \delta(\hs+\htm+\hu)\ \ \  , 
\eeq
where  $f_{a/p}(x,Q^2)$ and  $f_{b/p}(x,Q^2)$  are the PDFs for the
colliding partons $a$ and $b$ in the interacting protons 
as functions of momentum fraction $x$, at scale $Q$,
$\dd \sigma/ \dd\htm$ is the hard scattering cross section of the
partonic subprocess $ab \to cd$, and the FF, $D_{h/c}(z_c, Q'^2)$
gives the probability for parton $c$ to fragment into hadron $h$
with momentum fraction $z_c$ at scale $Q'$. We use the convention
that the parton-level Mandelstam variables are written with a `hat' (like 
$\htm$ above). The scales are fixed in the present work as $Q = p_T/2$
and $Q' = p_T/(2z_c)$.

In this investigation  we use leading order (LO) partonic cross sections,
together with LO PDFs (GRV)\cite{GRV92} and FFs (KKP)\cite{KKP}. This ensures 
the consistency of the calculation. An advantage of the GRV parameterization 
is that this fit uses data down to a rather small scale ($\approx 0.25$ GeV$^2$),
and thus provides PDFs with approximate validity for hard processes down to relatively
small transverse momenta, $p_T \gtrsim p_{T0} = 1$ GeV (with our scale 
fixing). It can be argued that, to the extent 
that LO and next-to-leading order (NLO) PDFs and FFs are fitted to the same 
data, they represent different parameterizations of the same nonperturbative 
information. However, going to NLO will reduce the scale dependence 
of pQCD calculations\cite{aurenche99}. A NLO study along the lines of 
the present work is in progress\cite{pzfbl02}.
An alternative approach is to use 
the ``Principle of Minimal Sensitivity''\cite{stevenson81}
to optimize these scales, as e.g. in Ref.~\cite{aurenche99}. However, fixing 
the scales is more convenient as a point of departure for $pA$ and $AA$ studies.

Since the fragmentation functions $D_{h/c}(z_c, Q'^2)$ are normally given for neutral 
hadrons or equivalently for the combination $h^\pm = (h^+ + h^-)/2$ (using isospin 
symmetry), assumptions need to be made to obtain the charge-separated FFs for
$\pi^+$, $\pi^-$, $K^+$, and $K^-$\cite{BKKI}. These assumptions are connected to 
the physical picture used to consider fragmentation from valence and sea quarks,
respectively. The simplest approximation corresponds to allowing a meson to 
fragment only from quarks that appear in it as valence contributions. We find
this prescription too restrictive, and use a 50-50 \% division of sea quark
contributions between positive and negative mesons in the present work 
($\sigma=0.5$),
unless indicated otherwise. The value of $\sigma$ hardly influences pion 
production, but has a more visible effect on kaons (in particular on $K^-$).
We illustrate this by also showing $\sigma=1$ results for comparison in figures
with $K/\pi$ ratios. The details of the approximation are discussed in the 
Appendix.  

The PDFs in eq. (\ref{hadX}) express the probability of finding a parton 
in the proton with longitudinal momentum fraction $x$, integrated over 
transverse momentum up to $k_T = Q$. Recent studies have considered 
parton distributions unintegrated over the transverse momentum $k_T$ of the
parton\cite{kimber00}. In a more phenomenological approach,
eq. (\ref{hadX}) can be generalized to incorporate intrinsic
transverse momentum by using a product assumption and
extending each integral over the parton 
distribution functions to $k_T$-space~\cite{Wang9798,Wong98},
\beq
\label{ktbroad}
\dd x \ f_{a/p}(x,Q^2) \rightarrow \dd x 
\ \dd ^2\!k_T\ g({\vec k}_T) \  f_{a/p}(x,Q^2) \ ,
\eeq
where $g({\vec k}_T)$ is the intrinsic transverse momentum distribution
of the relevant parton in the proton. We follow the phenomenological approach
in the present work, taking $g({\vec k}_T)$ to be a Gaussian:
\beq
\label{kTgauss}
g({\vec k}_T) \ = \f{1}{\pi \la k^2_T \ra}
        e^{-{k^2_T}/{\la k^2_T \ra}}    \,\,\, .
\eeq
Here $\langle k_T^2 \rangle$ is the 2-dimensional width of the $k_T$
distribution and it is related to the magnitude of the
average transverse momentum of one parton
as $\langle k_T^2 \rangle = 4 \langle k_T \rangle^2 /\pi$.
In order to regularize the singularity in the cross sections 
(associated with one of the Mandelstam variables approaching zero) 
we use the standard procedure of introducing a regulator mass 
$M=0.8$~GeV. As discussed e.g.
in  \cite{Wang9798}, the results in the 
addressed energy and transverse-momentum range are not sensitive to 
reasonable variations in this quantity. Kinematical details can be found in
Ref.s \cite{wang01;b,Wang9798,Wong98}.

The need for intrinsic transverse momentum in $pp$ collisions was 
investigated as soon as pQCD calculations were applied to
reproduce high-$p_T$ hadron production~\cite{owens87,sivers76}. 
An average intrinsic transverse momentum of
$\kav \sim 0.3-0.4$~GeV could be easily understood in terms of
the Heisenberg uncertainty relation for partons inside the proton.
However, a larger 
average transverse momentum of $\kav \sim 1$ GeV was extracted 
from jet-jet angular distributions (see e.g. \cite{E609}).
Recently, new theoretical efforts were mounted to
understand the physical origin of $\kav$~\cite{kimber00,Guo96,Lai98}.
Intrinsic transverse-momentum distributions have been utilized in the 
contexts of photon production\cite{ApaBep,plf00,Hust} and 
$J/\psi$ production\cite{sridhar98,zhuang01}. 
The description of deep inelastic 
scattering in the pomeron framework also appears to require 
a broad intrinsic $k_T$ distribution\cite{nikolaev99}.

To test the validity of the above approach, we calculated $\pi^0$, $\pi^+$ 
and $\pi^-$ production in $pp$ collisions in the energy range
20 GeV $\lesssim \sqrt{s} \lesssim$ 60 GeV, and compared the results 
to data from several independent experiments
\cite{antreasyan79,R806,CCRS,E605pi,E605pib,ISR}.
The calculations were performed using
the finite rapidity windows of the data. 
The Monte-Carlo integrals were carried out by the upgraded
VEGAS-routine~\cite{VEGAS}. 
As discussed above, the scales are fixed as $Q = p_T/2$ and $Q' = p_T/(2z_c)$. 
At each measured 
transverse momentum value ($p_T  \geq 2$ GeV) we determined the width of 
the transverse-momentum distribution, $\k2av$, necessary to fit the data point
at the given $p_T$, together with its error bar. The values of the extracted 
$\k2av$ are shown in Fig. 1 for $\pi^0$, and in Fig. 2. for $\pi^+$ (top panels) 
and $\pi^-$ (bottom panels) production. The error bars are determined from the 
experimental errors.

\begin{figure}
\vskip -0.2in
\epsfxsize=3.in
\epsfysize=3.in
\centerline{\epsffile{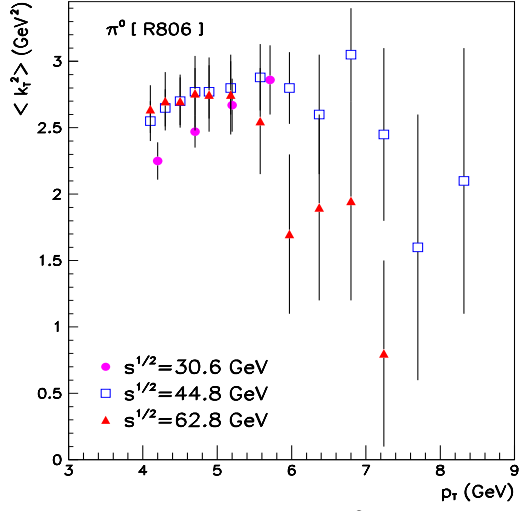}}
\vskip -0.05in
\caption[]{
\label{figure1}
The width of the transverse momentum distribution, $\k2av$, necessary 
in eq. (\ref{kTgauss}) to fit the measured value of the spectrum at
the given $p_T$, point by point, for $\pi^0$ production
in $pp$ collisions at three energies. The data are from Ref.~\cite{R806}.
}
\end{figure}

\begin{figure}
\vskip -0.6in
\epsfxsize=5.in
\epsfysize=4.5in
\centerline{\epsffile{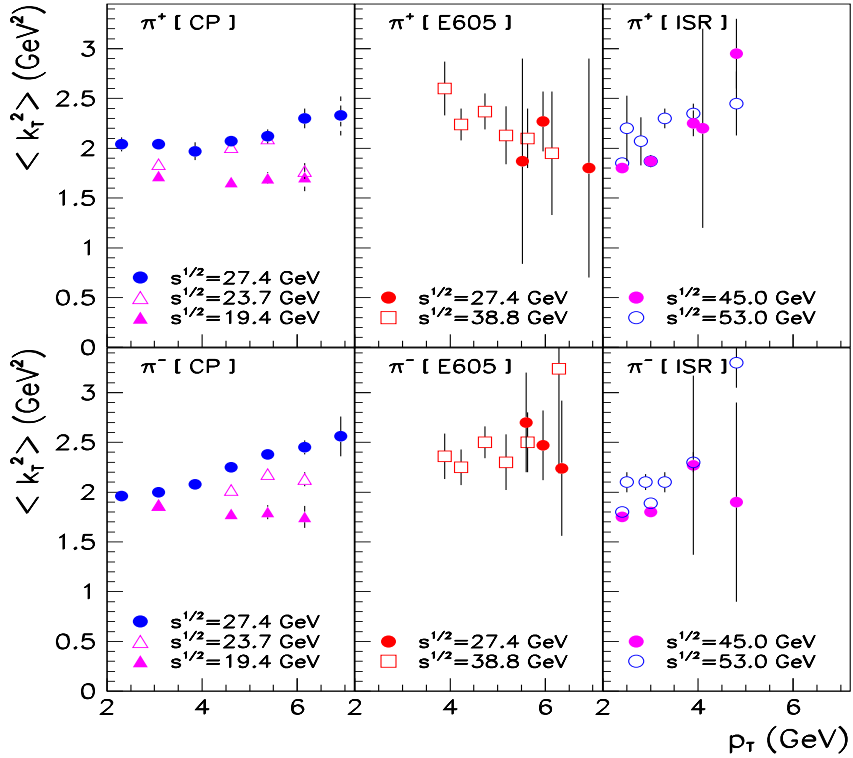}}
\vskip -0.05in
\caption[]{
\label{figure2}
The width of the transverse momentum distribution, $\k2av$, necessary 
in eq. (\ref{kTgauss}) to fit the measured value of the spectrum at
the given value of $p_T$, point by point, for $\pi^+$ (top panels)
and $\pi^-$ (bottom panels) production in $pp$ collisions at 
several energies. The data are from Ref.s 
\cite{antreasyan79,E605pi,E605pib,ISR}.
}
\end{figure}

It is clear from Fig.-s 1 and 2 that the description of the data in terms of 
a pQCD calculation augmented by a Gaussian transverse momentum distribution
makes sense only in a limited $p_T$ window in the energy range 
20 GeV $\lesssim \sqrt{s} \lesssim$ 60 GeV. In addition to the requirement 
of {\it hard} particle production ($p_T \geq p_{T0}$) imposed for the applicability
of pQCD, in most data sets there appears to be an upper limit in transverse momentum, 
to which $\k2av$ can be extracted with reasonable accuracy. In broad terms, 
the procedure may be considered sensible for, say 
2 GeV $\lesssim p_T \lesssim$ 7 GeV, 
depending on the details of the experiments. Within this window, 
one may reasonably extract an approximately constant (energy-dependent)
$\k2av$ for 3 GeV $\lesssim p_T \lesssim$ 6 GeV  in the full energy range.
(Note that the lower limit of the considered $p_T$ interval can be lowered with
increasing c.m. energy.) At higher transverse 
momenta (and fixed $\sqrt{s}$)
the effect of intrinsic $k_T$ becomes less important, and thus
it is not surprising that, 
using the error bars on the data,
the width $\k2av$ acquires a large uncertainty.
Although 3 GeV $\lesssim p_T \lesssim$ 6 GeV is a rather 
narrow transverse-momentum interval, the Cronin effect in $pA$ collisions,
to be addressed in the next subsection,
is most prominent in this range. This motivated us to use an energy-dependent,
$p_T$-independent $\k2av$ in what follows, as an approximation.

Concentrating on the above interval in $p_T$, we next determined 
the $p_T$-independent $\k2av$
which best fits the pion production data in $pp$ collisions as a function of 
c.m. energy. In a slight departure from our earlier work\cite{plf00}, we fitted the data 
minimizing the standard $\chi^2=\sum (Data-Theory)^2/(Exp. uncertainty)^2$
to obtain an optimal $\k2av$ for each energy. 
Fig. 3 summarizes these results. Data are from a range of independent 
experiments\cite{antreasyan79,R806,CCRS,E605pi,E605pib,ISR}.
%,R807,E704pi,NA24pi,R110pi,WA70

\begin{figure}
\vskip -0.4in
\epsfxsize=5.5in
\epsfysize=6in
\centerline{\epsffile{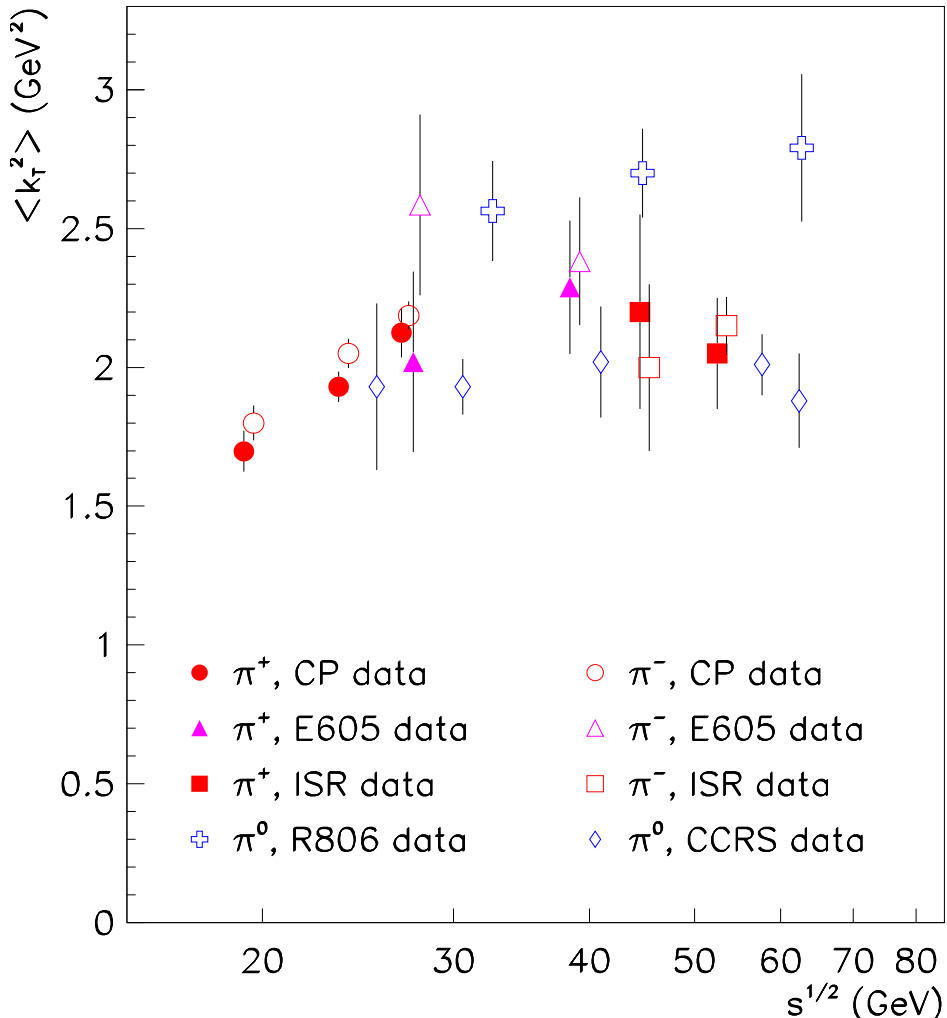}}
\vskip -0.05in
\caption[]{
\label{figure3}
The best fit values of $\k2av$ in $p p \rightarrow \pi X$ 
\cite{antreasyan79,R806,CCRS,E605pi,E605pib,ISR}
%,E704pi,NA24pi,R110pi,WA70
reactions. Where larger error bars would overlap, the $\pi^-$ point
has been shifted slightly to the right for better visibility.
}
\end{figure}

We conclude from this figure that the value of $\k2av$ for hard pion production 
shows an increasing trend 
as a function of $\sqrt{s}$ from 20 to 60 GeV (and may be 
leveling off towards the high-energy end of the interval). The numerical values
are significantly larger than expected solely on the basis of the Heisenberg
uncertainty principle. It needs to be kept in mind that these values are obtained 
in the framework of a LO calculation. At NLO, smaller transverse width should be
required for the best fit of the data 
(see some results in Ref. \cite{bflpz01}). 
At present, we focus on the utility of 
this parameterization to describe the $pp$ data and view Fig. 3 as a basis for 
addressing $pA$ and $AA$ collisions at the same level of pQCD.  

As an example to illustrate the degree of accuracy of the description in the $pp$ 
sector, Fig. 4 compares calculated $\pi^+$ and $\pi^-$ 
spectra and $\pi^-/\pi^+$ ratios to the data\cite{antreasyan79} at the c.m. energies 
$\sqrt{s} =$ 19.4, 23.8, and 27.4 GeV, respectively, obtained with the
values of $\k2av$, indicated in the top panels. 
We find that the data/theory and $\pi^-/\pi^+$ ratios are well reproduced 
for 2 GeV $\lesssim p_T \lesssim$ 6~GeV. Based on this and similar 
examples, we believe that hard pion 
production in $pp$ collisions is reasonably under control at the present
level of calculation. With the parameterization of the transverse momentum 
distribution we separated any pQCD uncertainties from the nuclear effects,
which we address in the next subsection.     
 
\begin{figure}
\vskip -0.2in
\epsfxsize=5.5in
\epsfysize=6in
\centerline{\epsffile{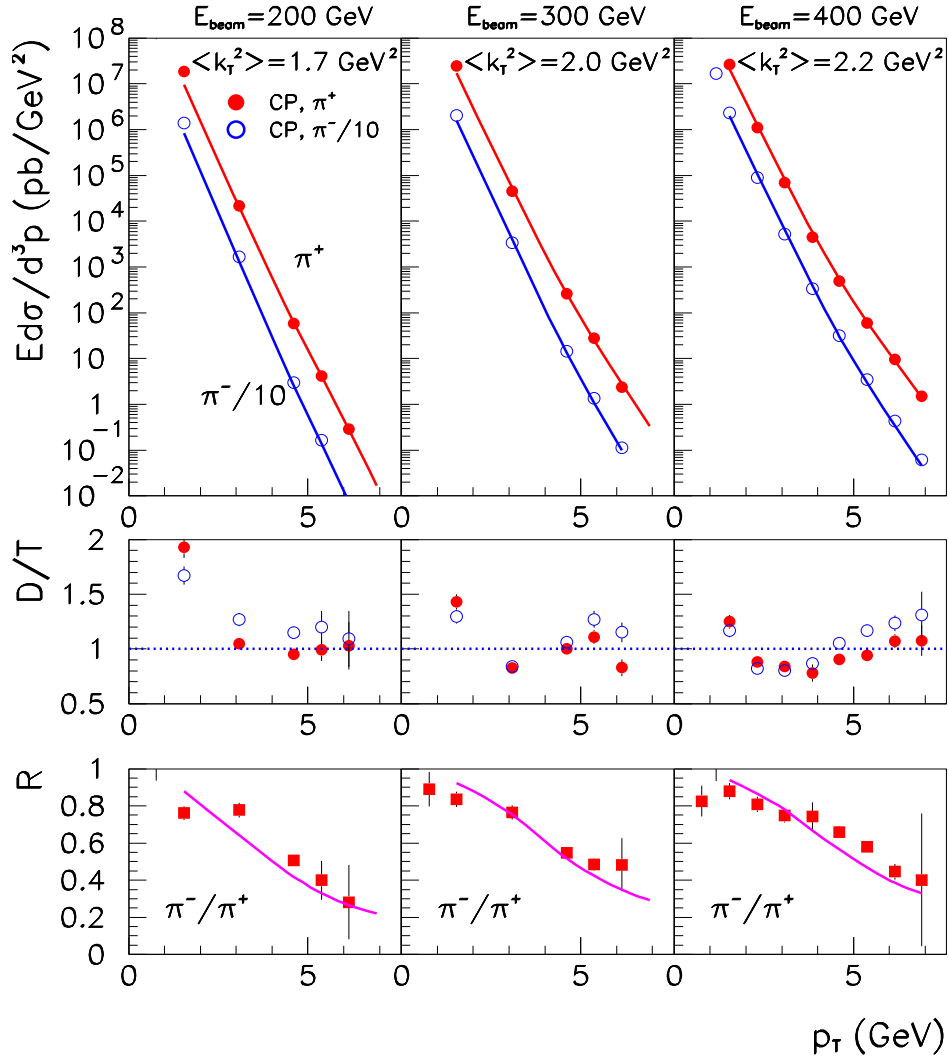}}
\vskip -0.05in
\caption[]{
\label{figure4}
Invariant cross section of $\pi^+$ and $\pi^-$ production from $pp$
collisions (top row), D/T: data/theory ratios (middle row),
and R: $\pi^-/\pi^+$ ratios (bottom row)
as functions of transverse momentum at c.m. energies  
$\sqrt{s} =$ 19.4, 23.8, and 27.4 GeV. Data are from \cite{antreasyan79}. 
The calculations are carried out in LO, as explained in the text.
}
\end{figure}

\subsection{Hard pions from proton-nucleus collisions}
\label{sec_nu}

Interest in the nuclear dependence of hard particle production was 
triggered by the discovery of the Cronin effect~\cite{cronin75,antreasyan79},
and revived with the study of collisions of $\alpha$ particles at the 
CERN Interacting Storage Ring (ISR)\cite{kuhn76,sukhatme82,lev83}.
As discussed in the introduction, the nuclear enhancement of hard
pion and kaon production cross sections is not only interesting in itself,
but we also need to understand these phenomena before we can move to 
the description of $AA$ collisions at RHIC energies. 

The standard framework for addressing high-energy $pA$ collisions is
provided by the Glauber model\cite{glauber59,gribov}. It should be 
kept in mind, however, that the Glauber model was originally developed with 
nucleons as elementary constituents, neglecting coherent scattering from 
several nucleons. Therefore, it is not surprising that 
refinements to the standard Glauber picture are proposed based on the 
underlying quark-gluon structure\cite{peitzmann99,gale99,hufner00}. As
such modifications are not yet generally accepted or organized into a consistent 
picture, we continue using the standard Glauber 
description in the present work. 

The key observation for the explanation of the Cronin effect is that
in $pA$ collisions, in addition to the hard parton scattering, the
incoming and outgoing partons may suffer additional interactions in
the presence of the nuclear medium\cite{wang01;b,Wang9798,Wong98}. 
An effective way to summarize 
these additional interactions is in terms of an enhancement of the 
width of the transverse momentum distribution above the value in $pp$
collisions, shown in Fig. 3. The extra
contribution to the width due to the nuclear environment
can be related to the number of nucleon-nucleon collisions 
in the medium. To characterize the $\k2av$ enhancement, we write the
width of the transverse momentum distribution of the partons
in the incoming proton as
\beq
\label{ktbroadpA}
\k2av_{pA} = \k2av_{pp} + C \cdot h_{pA}(b) \ . 
\eeq
Here $\k2av_{pp}$ is the width of the transverse momentum distribution 
of partons in $pp$ collisions from Subsection \ref{sec_kT} (also denoted simply by
$\k2av$), $h_{pA}(b)$ describes the number of {\it effective} 
nucleon-nucleon (NN) collisions at impact parameter $b$
which impart an average transverse momentum squared $C$. In $pA$
reactions, where one of the partons participating in the hard collision
originates in a nucleon with additional NN collisions, we will use the
$pp$ width from Fig. 3 for one of the colliding partons and the 
enhanced width (\ref{ktbroadpA}) for the other. The function 
$h_{pA}(b)$ will be discussed shortly.

According to the Glauber picture, the hard pion production cross section 
from $pA$ reactions can be written as an integral over impact parameter $b$:
\beq
\label{pAX}
  E_{\pi}\f{\dd \sigma_{\pi}^{pA}}{ \dd ^3p} =
       \int \dd ^2b \,\, t_A(b)\,\, E_{\pi} \, \f{\dd \sigma_{\pi}^{pp}(\k2av_{pA},\k2av_{pp})}
{\dd ^3p}  
\,\,\, ,
\eeq
where the proton-proton cross section on the right hand side represents 
the cross section from eq. (\ref{hadX}) with the transverse extension 
as given by eq.-s (\ref{ktbroad}) and (\ref{kTgauss}), but with the 
widths of these distributions as indicated. Here $t_A(b) = \int \dd z \, \rho(b,z)$ 
is the nuclear thickness function (in terms of the density distribution $\rho$)
normalized as $\int \dd ^2b \, t_A(b) = A$.
Furthermore, it is well-known that the PDFs are modified in the nuclear 
environment (``shadowing'')\cite{wang91,eskola99}. We approximately include 
shadowing and the isospin asymmetry of heavy nuclei into the nuclear 
PDFs by considering the average nuclear dependence and using a 
scale independent parameterization of the shadowing function $S_{a/A}(x)$ 
adopted from Ref. \cite{wang91}:
\beq
\label{shadow}
f_{a/A}(x,Q^2) = S_{a/A}(x) \left[\frac{Z}{A} f_{a/p}(x,Q^2) + \left(1-\frac{Z}{A}\right)
  f_{a/n}(x,Q^2) \right]   \,\,\,\,  ,
\eeq
where $f_{a/n}(x,Q^2)$ is the PDF for the neutron.

The effectivity function $h_{pA}(b)$ can be written in terms of the 
number of collisions suffered by the incoming proton in the target 
nucleus, $\nu_A(b) = \sigma_{NN} t_A(b)$, where $\sigma_{NN}$ is
the inelastic nucleon-nucleon cross section. Two extreme prescriptions were used for
$h_{pA}(b)$ earlier: (i) $h_{pA}^{all}(b) = \nu_A(b)-1$, corresponding to all NN
collisions, except the hard interaction producing the parton to fragment
into the observed pion, contributing to the $\k2av$ enhancement\cite{wang01;b}, and
(ii) a `saturated' prescription, where $h_{pA}^{sat}(b)$ was equated to a maximum 
value of unity whenever $\nu_A(b) \geq  \nu_m=2$ (i.e. it was assumed that only
one associated NN collision is responsible for the $\k2av$ enhancement, 
further collisions are not important in this regard)\cite{plf00}. 
In the present work, we examine the dependence of the results on the possible choices
between these limits, denoted by $\nu_m=\infty$ and $\nu_m=2$, respectively.
Since even in a heavy nucleus $\nu_A(b)$ does not exceed $\approx 6$,
we carried out explicit calculations for $2 \leq \nu_m \leq 5$ and without any cut 
($\nu_m=\infty$). The coefficient $C$ in eq. (\ref{ktbroadpA}) best 
describing the available $pA$ data\cite{antreasyan79} 
for the different values of $\nu_m$ is shown at three energies
in Fig.~5 as a function of $p_T$ for Be, Ti, and W target nuclei.

\begin{figure}
\epsfxsize=5.5in
\epsfysize=6in
\centerline{\epsffile{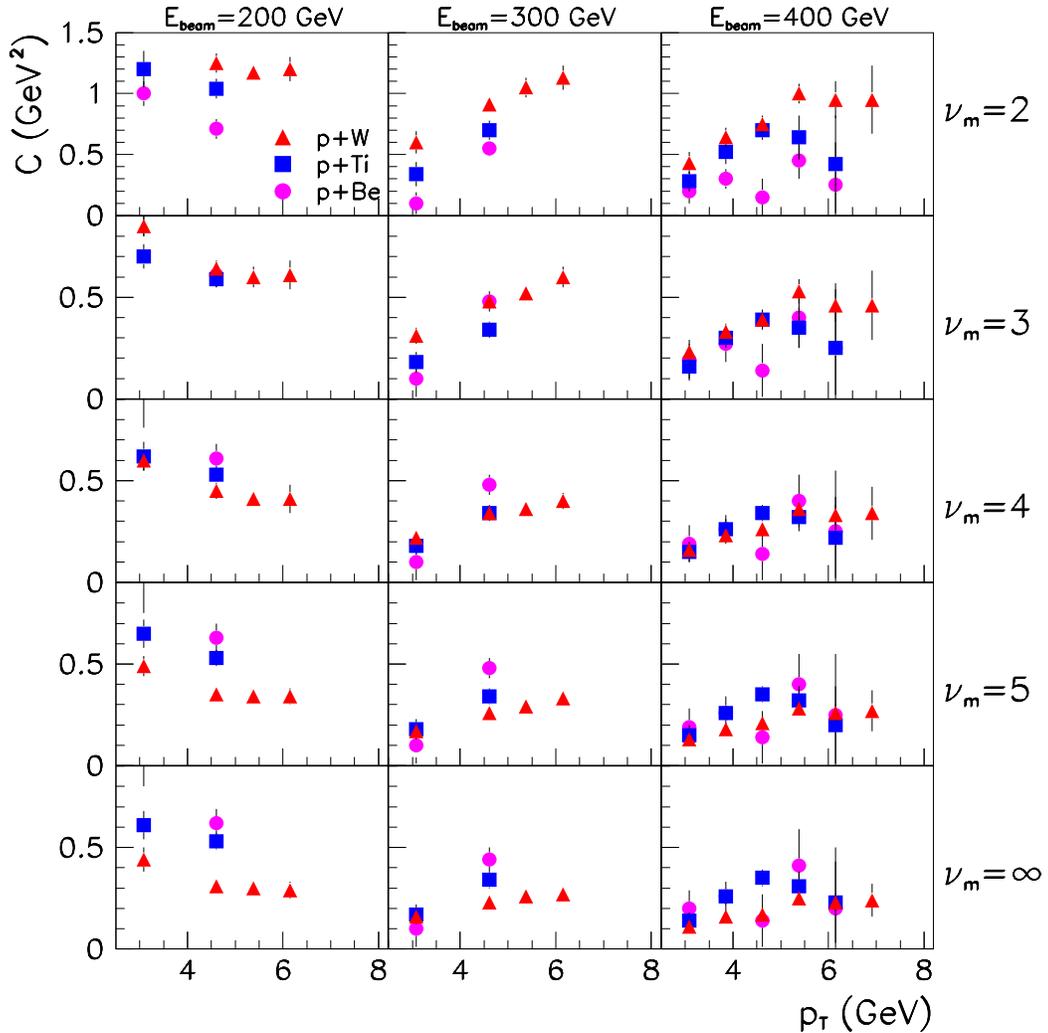}}
\vskip -0.05in
\caption[]{
 \label{figure5}
The best fit values of $C$ in eq. (\ref{ktbroadpA}) for $Be, Ti$, and $W$ targets
at three different energies as a function of $p_T$ and $\nu_m$. The fitted
data are from Ref. \cite{antreasyan79}.
}
\end{figure}

For an appealing physical interpretation, the coefficient $C$ is expected
to be approximately independent of $p_T$, of the target used,
and probably of beam energy (at least in the energy range studied in Fig. \ref{figure5}). 
Based on these requirements, we select $\nu_m=4$ as the value that 
produces results for $C$ closest to the ideal physical picture. In what 
follows, we fix $\nu_m=4$, where the associated value of $C$ is  
$C=0.40 \pm 0.05$~GeV$^2$.
This value of $C$ is significantly smaller than what we needed to use in
connection to $\nu_m=2$ earlier ($C^{sat}$, see Ref. \cite{plf00}), and 
somewhat larger than would be needed in the case of $\nu_m=\infty$. 
This can be understood based on the compensating effects of $h_{pA}(b)$ and the 
value of $C$ in eq.~(\ref{ktbroadpA}).  

It is interesting to note that a similar value for the
number of effective inelastic ("semi-hard") collisions 
was obtained examining nuclear stopping
in $p+Be,Cu,Au$ collisions at lower energies 
($E_{beam}=$12 and 18 AGeV)\cite{cole,Chem,Cole01}. The leading proton appeared
in these experiments not to loose more energy 
after 2 -- 3 collisions (measured via 'grey' tracks from the target).
The physics of this phenomenon was connected to the constituent
quark picture of the proton, i.e. it was suggested 
that the state resulting after the inelastic excitation 
of the three constituent quarks 
no longer interacts with a high probability.
Further experiments are needed to verify this 
or a similar physical interpretation. Such an effort is  
in progress at AGS and at CERN SPS (see Ref. \cite{Cole01,Veres01}).
Our finding of $\nu_m=4$, which means $\sim 3$ effective semi-hard collisions,
overlaps with the above result and points to a similar mechanism, even at
higher energies. At this point, the present study lends phenomenological 
support to the explanation advanced in Ref. \cite{cole,Chem,Cole01}.
Our model is also consistent with the idea proposed by Dokshitzer \cite{doksh01}
about the proton being unable to accept more than a characteristic
momentum transfer of $Q \approx 1 - 1.2$ GeV. This idea implies the existence
of a limiting value of $\langle k_T^2 \rangle$ for the proton, which can be
accumulated in one or several semi-hard interactions. 
Further data are needed at CERN SPS and RHIC energies for a 
more definitive statement. The planned $pA$ collision program
at RHIC\cite{pARHIC} may play a very important role in this regard.

To illustrate the quality of agreement with the data achieved with the 
above parameterization, Fig. 6 displays $\pi^+$ spectra (upper panels) 
and $\pi^-/\pi^+$ ratios (lower panels) as functions of $p_T$,
compared to the same $pA$ data\cite{antreasyan79} as used to determine $C$. The 
parameters are fixed as indicated in the top panels: $\nu_m=4$, $C=0.4$ GeV$^2$, 
and the energy-dependent $\k2av_{pp}$ from Fig. \ref{figure3}. There is no
observable difference in the $\pi^-/\pi^+$ ratios between the different targets
in either the data or the calculated results.  

\begin{figure}
\vskip -0.2in
\epsfxsize=5.5in
\epsfysize=6in
\centerline{\epsffile{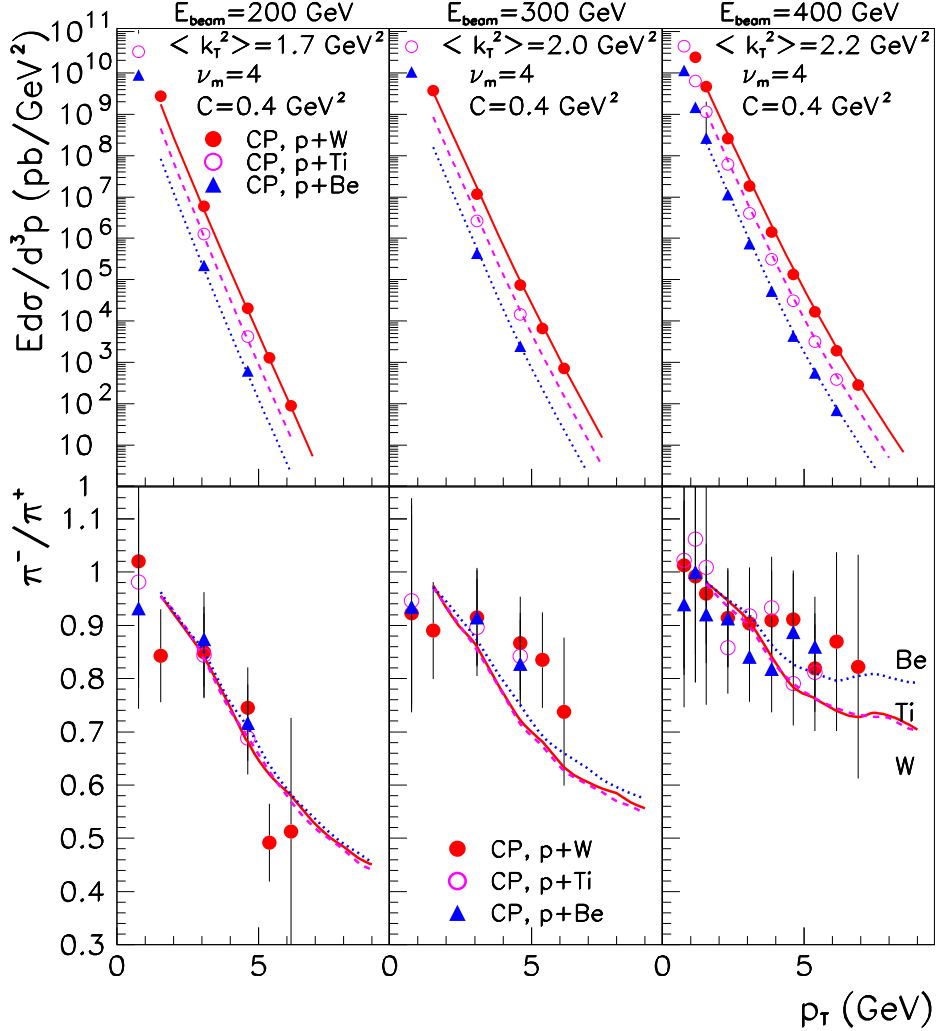}}
\vskip -0.05in
\caption[]{
\label{figure6}
$\pi^+$ spectra and $\pi^-/\pi^+$ ratios from $pA$ collisions 
for $A=Be,Ti,W$ as functions of transverse momentum 
at c.m.~energies  $\sqrt{s} =$ 19.4, 23.8, and 27.4 GeV. Data are
from \cite{antreasyan79}. 
}
\end{figure}

\begin{figure}
\vskip -0.4in
\epsfxsize=4.5in
\epsfysize=5in
\centerline{\epsffile{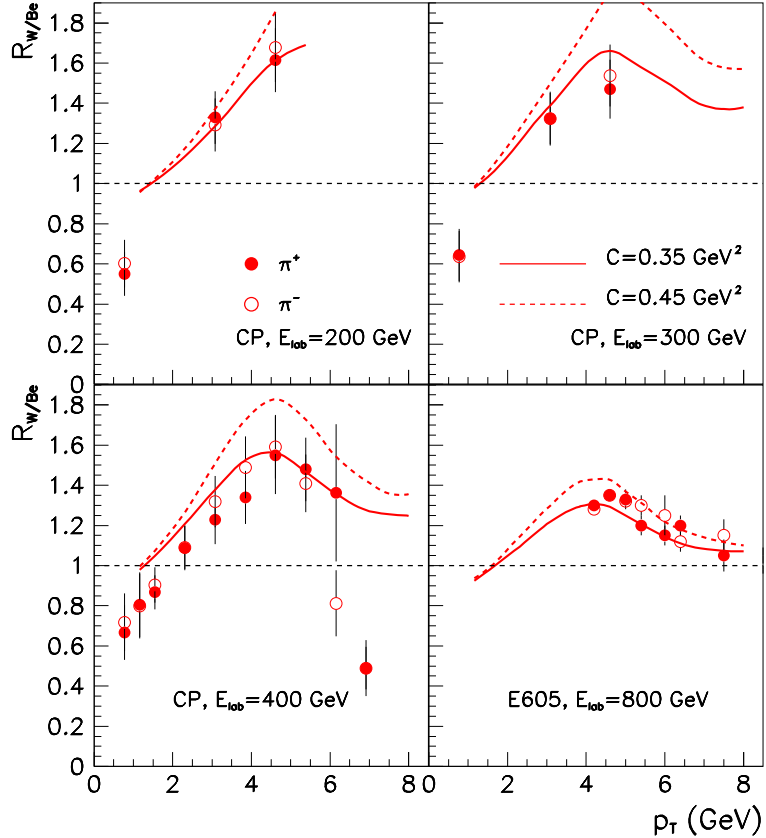}}
\vskip -0.05in
\caption[]{
\label{figure7}
Normalized $W/Be$ charged pion cross section ratios at four energies. The 
calculations are carried out with the parameters fixed earlier as discussed 
in the text. The data are from Refs. \cite{antreasyan79,E605pi,E605pib}. 
The deviation from unity represents the Cronin enhancement.
}
\end{figure}

The Cronin enhancement can be presented in the form of normalized (by mass number $A$) 
$W/Be$ cross section ratios\cite{antreasyan79,E605pib,brown96}.
The data for $\pi^+$ and $\pi^-$ are shown as a function of $p_T$ for the 
above energies and for $E_{lab}=800$~GeV~\cite{E605pib}
in Fig. \ref{figure7}, together with the results of calculations with $C=0.35$ GeV$^2$ 
(solid line) and with $C=0.45$ GeV$^2$ (dashed line). The band between these two curves 
represents the uncertainty in our calculation associated with the extraction 
of the coefficient $C$ (see Fig. \ref{figure5}). 

In the absence of the Cronin effect, these ratios would give a value of identically 1.
The significant deviation of the data from unity is therefore a clear confirmation of 
the nuclear enhancement in the 2 GeV $\lesssim p_T \lesssim$ 6 GeV transverse momentum
window. At lower $p_T$, where absorption effects set in, the data indicate a value of
the ratio smaller than one. However, we do not trust our pQCD calculation below
$p_T \approx 2$ GeV, and thus do not show calculated results for low transverse momenta. 
At high transverse momenta, where the effects of intrinsic $k_T$ become unimportant 
as mentioned earlier, the ratio converges to unity in both the data and the calculations.  

\subsection{Nucleus-nucleus collisions}
\label{sec_AApi}

In nucleus-nucleus reactions\footnote{These are denoted 
by $AA$ in the text, even 
though not all studied systems are symmetric.}, where both partons entering the hard 
collision originate in nucleons with additional 
semi-hard collisions, we apply the
enhanced width of the parton distribution (\ref{ktbroadpA}) for both
initial partons. Thus,
\beq
\label{ABX}
  E_{\pi}\f{\dd \sigma_{\pi}^{AB}}{\dd ^3p} =
       \int \dd ^2b \, \dd ^2r \,\, t_A(r) \,\, t_B(|\vec b - \vec r|) \,
E_{\pi} \, \f{\dd \sigma_{\pi}^{pp}(\k2av_{pA},\k2av_{pB})}{\dd ^3p}  
\,\,\, ,
\eeq
where the proton-proton cross section on the right hand side represents 
the cross section from eq. (\ref{hadX}) with the transverse extension 
as given by eq.-s (\ref{ktbroad}) and (\ref{kTgauss}), with the enhanced 
widths of these distributions given by (\ref{ktbroadpA}). 
We emphasize that the parameter values used to
describe hard pion production in $AA$ reactions are identical to
the ones used earlier in this Section for the best description
of pion production from $pp$ and $pA$ reactions. In other words,
the values of $\k2av_{pp}=$1.6 and 1.7~GeV$^2$, respectively,
are taken from Fig. \ref{figure3} (with
a relatively small uncertainty of $\approx \pm 0.1$ GeV 
at $\sqrt{s} \sim 17-20$ GeV), and $\nu_m = 4$ and $C = 0.4$~GeV$^2$ 
are fixed.    
    
\begin{figure}
\vskip -0.2in
\epsfxsize=5.5in
\epsfysize=6in
\centerline{\epsffile{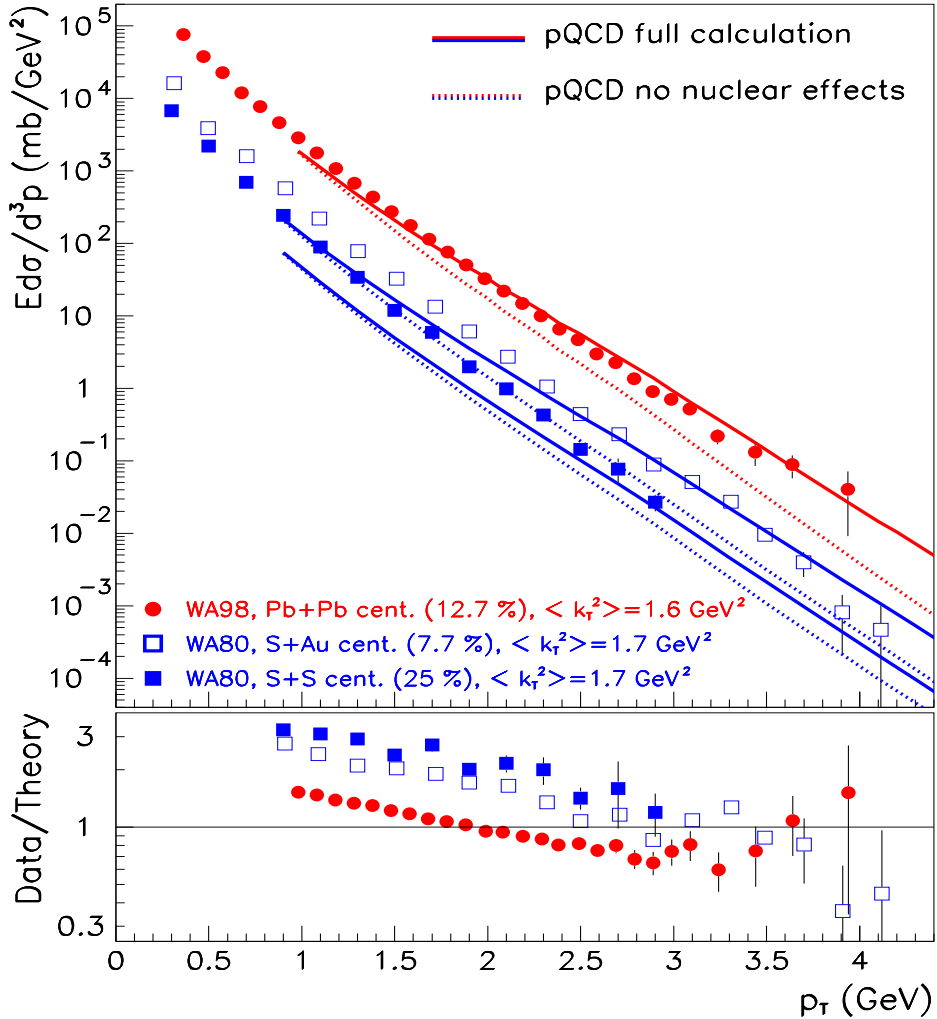}}
\vskip -0.05in
\caption[]{
\label{figure7a}
Neutral pion production compared to data from
SPS experiments WA80\cite{WA80} and WA98\cite{WA98} for central collisions. 
Calculated invariant cross sections with the nuclear effects turned off
(dotted lines) and with the complete model (full lines) are displayed
on top together with the corresponding data, data/theory ratios 
are shown on the bottom (with the full calculation). 
The parameters are fixed by $pp$ and $pA$ information as discussed in the text.
}
\end{figure}

The results of these calculations for central $\pi^0$ spectra 
are compared to the WA80\cite{WA80}
and WA98\cite{WA98} data for $S+S$, $S+Au$, and $Pb+Pb$ collisions
in Fig. \ref{figure7a} as a function of 
 transverse momentum. Spectra are shown in the top panel,
data/theory ratios are displayed at the bottom. The dotted lines represent the 
results of the pQCD calculation with nuclear effects
(shadowing and multiscattering) turned off;
solid lines correspond to the full calculation. When comparing 
to the data, it should be kept in mind that the calculation 
is intended for {\em hard} particle production, and it is not
expected to describe the data below $p_T \approx$ 2 GeV. 
At low $p_T$, where soft processes become important, the 
pQCD model underestimates the  $S + S$ data, as expected.
The calculation without nuclear effects (dotted lines) 
underestimates the data in the entire measured $p_T$ range.
The reproduction of the $S+S$ and $S+Au$ data can be 
considered reasonable for $p_T \gtrsim $ 2 GeV, while 
the $Pb+Pb$ calculation overestimates the data in the same transverse momentum 
window by upto 40\%. This may be taken as a hint that  
an additional mechanism is at work in the nuclear medium,
which acts to reduce the calculated cross sections. 
Jet quenching\cite{plumer,baier98,gyulassy00} is a potential candidate 
for the physics not included in the present model.

%\newpage
\section{Hard kaon production at c.m. energies below $\sim 60$ GeV}
\label{sec_K}     

In this Section we turn to charged kaon production in the same c.m. energy 
region, using the parameter values fixed in Section \ref{sec_pi}. 
We are interested in the reproduction of kaon data by including the effects of
multiscattering and shadowing. Most data are available as kaon to pion ratios;
we will display the calculated results in the same manner.
Furthermore, as mentioned above, $K^-$ production is particularly
sensitive to assumptions made in connection with sea quark fragmentation,
discussed in the Appendix. Our standard computations are carried out with
the value of $\sigma=0.5$ for the parameter governing the distribution of 
sea quarks, but as a reminder to this dependence and to illustrate the
sensitivity, we also show calculated results with $\sigma=1$ in this Section.

\subsection{Charged kaons from proton-proton collisions}
\label{sec_ppK}

\begin{figure}
\vskip -0.2in
\epsfxsize=5.in
\epsfysize=5.5in
\centerline{\epsffile{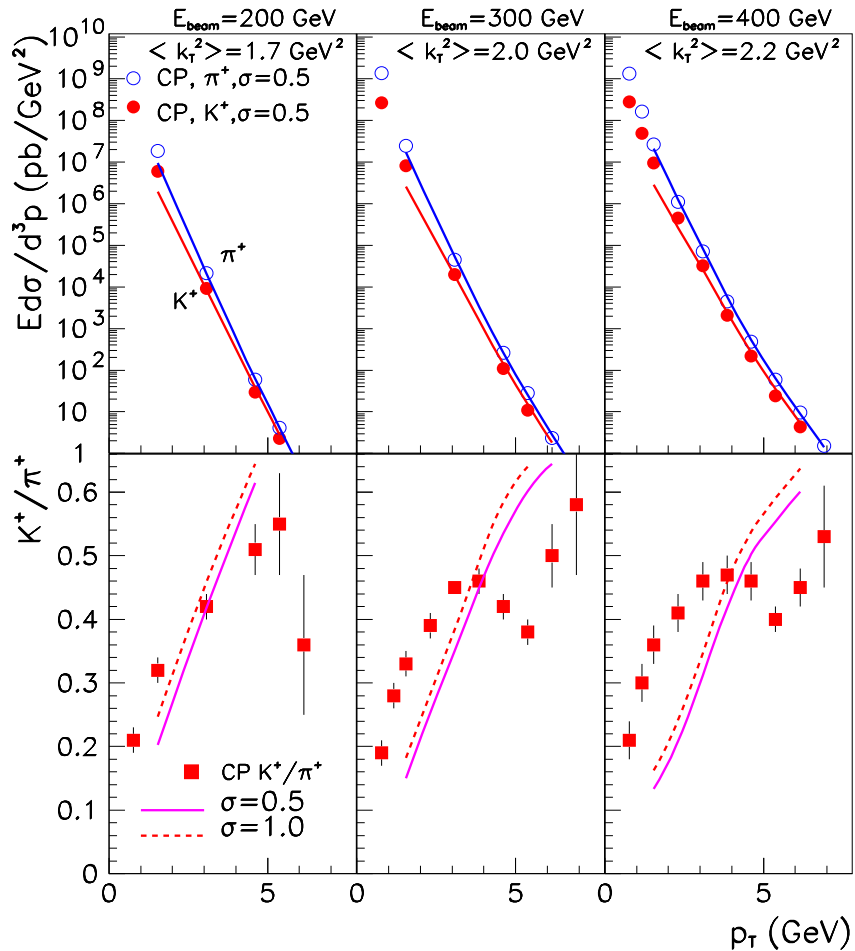}}
\vskip -0.05in
\caption[]{
 \label{figure8}
Spectra of $\pi^+$ and $K^+$ mesons as functions of transverse momentum from $pp$ collisions
at c.m. energies  $\sqrt{s} =$ 19.4, 23.8, and 27.4 GeV (top panels) and $K^+/\pi^+$ ratios
calculated with $\sigma=0.5$ (solid) and $\sigma=1$ (dashed) as explained in the text
(bottom panels). Data are from \cite{antreasyan79}.
}
\end{figure}

\begin{figure}
%\vskip -0.4in
\epsfxsize=5.5in
\epsfysize=6in
\centerline{\epsffile{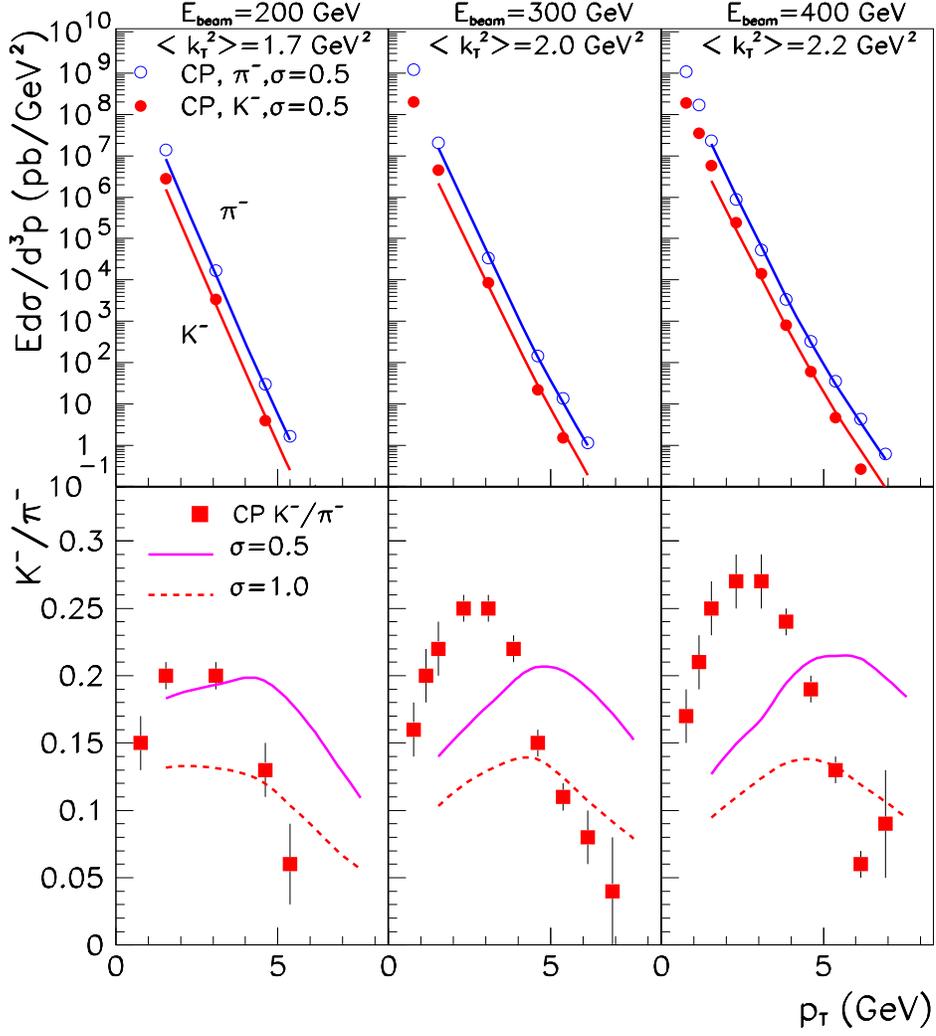}}
\vskip -0.05in
\caption[]{
 \label{figure9}
Spectra of $\pi^-$ and $K^-$ mesons as functions of transverse momentum from $pp$ collisions 
at c.m. energies  $\sqrt{s} =$ 19.4, 23.8, and 27.4 GeV (top panels) and $K^-/\pi^-$ ratios
calculated with $\sigma=0.5$ (solid) and $\sigma=1$ (dashed) as explained in the text
(bottom panels). Data are from \cite{antreasyan79}.
}
\end{figure}

First we look at kaon spectra and $K/\pi$ ratios for positive 
and negative mesons from $pp$ collisions to check if the 
parameterization introduced for pions is also applicable in this case.
Our results are compared to the data\cite{antreasyan79} 
in Fig.~\ref{figure8} for $K^+$ and in Fig. \ref{figure9} for $K^-$ 
at the c.m. energies  $\sqrt{s} =$ 19.4, 23.8, and 27.4 GeV. In the 
top panels the appropriate pion spectra are also shown
as a reference. The kaon cross sections displayed here
are calculated with the same value of $\sigma=0.5$. 
The data/theory ratios show an agreement similar to that   
seen in Fig. \ref{figure4} for pions.
In the bottom panels 
presenting the $K/\pi$ ratios, full lines correspond to the standard $\sigma=0.5$,
dashed lines depict the calculated results with $\sigma=1$. 

\begin{figure}
%\vskip -0.4in
\epsfxsize=5.5in
\epsfysize=6in
\centerline{\epsffile{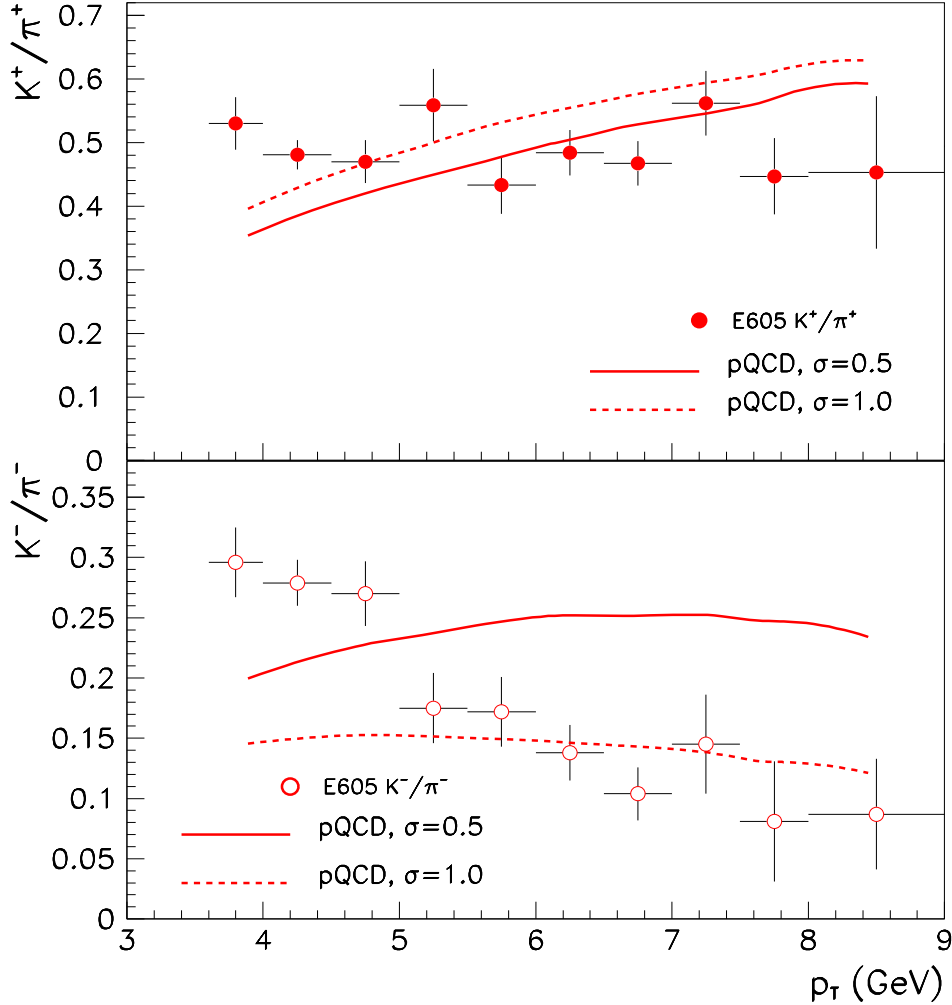}}
\vskip -0.05in
\caption[]{
 \label{figure10}
Ratios of $K^+/\pi^+$ (top) and $K^-/\pi^-$ (bottom) at $E_{beam}=800$ GeV,  
calculated with $\sigma=0.5$ (solid) and $\sigma=1$ (dashed).
Data are from \cite{E605pi,E605pib}.
}
\end{figure}

The reproduction of the kaon data from $pp$ collisions
can be considered fair, except that in the $K/\pi$
ratios the apparent structure of the $K^+/\pi^+$ ratios is not reproduced,
and the peak in the $K^-/\pi^-$ ratios is shifted 
to higher transverse momenta and is less pronounced than in the 
data. It can also be seen that the distribution of sea quarks for the fragmentation 
of charged mesons, while not too important 
for the $K^+/\pi^+$ ratio, makes a significant difference for $K^-/\pi^-$, with 
$\sigma=0.5$ describing these ratios better than $\sigma=1$ for $p_T \lesssim 5$ GeV.
If we focus on $p_T > 5$ GeV, the opposite conclusion may be drawn.
The sensitivity of the negative ratios to $\sigma$ is also larger in the 
E605 experimental $K/\pi$ ratios shown in Fig. \ref{figure10}. Here, 
measurements extend to higher values of $p_T$, and $\sigma=1$
gives a better fit to the $K^-/\pi^-$ ratios for $p_T \gtrsim 5$ GeV, similarly to the 
$K^-/\pi^-$ ratios of Ref. \cite{antreasyan79}. 
This may be taken as a hint that our prescription for the 
fragmentation contribution of sea quarks is too schematic, and $\sigma$ should 
be taken $p_T$-dependent in a more realistic calculation.

\subsection{Kaons from proton-nucleus collisions}
\label{sec_pAK}

Next we examine the predictive power of our model for charged kaon production
from $pA$ collisions at c.m. energies  $\sqrt{s} =$ 19.4, 23.8, and 27.4 GeV,
with the parameters fixed as above. In other words, $\k2av$ is taken from 
Fig. \ref{figure3} as in Subsection \ref{sec_ppK}, and the enhancement of 
the transverse momentum width is given by eq. (\ref{ktbroadpA}) using $\nu_m=4$
and $C=0.4$ GeV$^2$.

\begin{figure}
%\vskip -0.4in
\epsfxsize=5.5in
\epsfysize=6in
\centerline{\epsffile{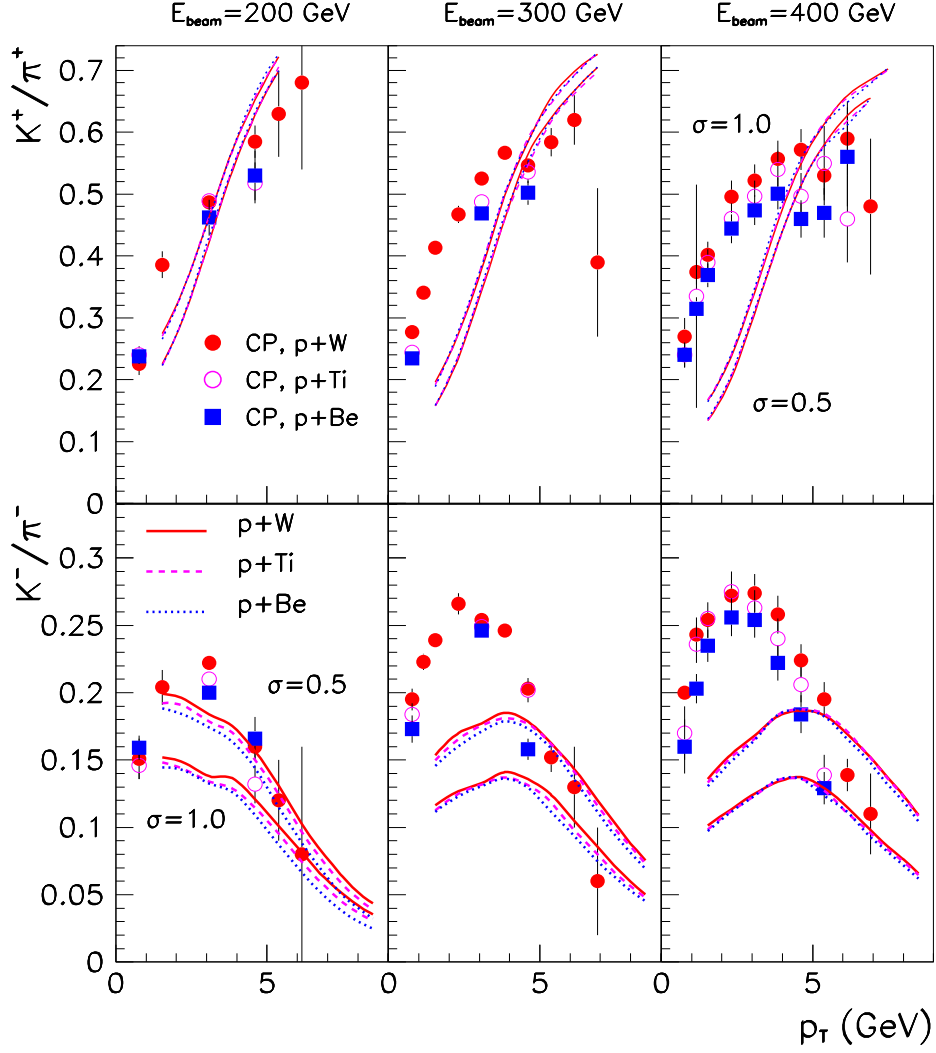}}
\vskip -0.05in
\caption[]{
 \label{figure11}
$K^+/\pi^+$ (top) and $K^-/\pi^-$ (bottom) ratios for three nuclei
at c.m. energies  $\sqrt{s} =$ 19.4, 23.8, and 27.4 GeV. The two bands of
calculations correspond to $\sigma=0.5$ and $\sigma=1$, respectively.
Data are from \cite{antreasyan79}.
}
\end{figure}

The data are available as $K/\pi$ ratios, eliminating experimental normalization
issues\cite{antreasyan79}. In Fig. \ref{figure11} we compare the results of our
calculations to these data for $Be, Ti$, and $W$ targets. $K^+/\pi^+$ ratios are
shown in the top panels, while $K^-/\pi^-$ ratios are displayed on the bottom.
We present calculations with both $\sigma=0.5$ and $\sigma=1$. The general tendency
of these results and the quality of the agreement between the data and the 
calculations is very similar to that observed in $pp$ collisions. The deviations
increase with energy, but the calculations seem to reflect the gross features of the 
data. The kaon to pion ratio is not sensitive to the target size. 
This means that the nuclear enhancement is similar to pions and kaons.

\subsection{Predictions for kaons from nucleus-nucleus collisions}
\label{sec_AAK}

\begin{figure}
%\vskip -0.4in
\epsfxsize=5.5in
\epsfysize=6in
\centerline{\epsffile{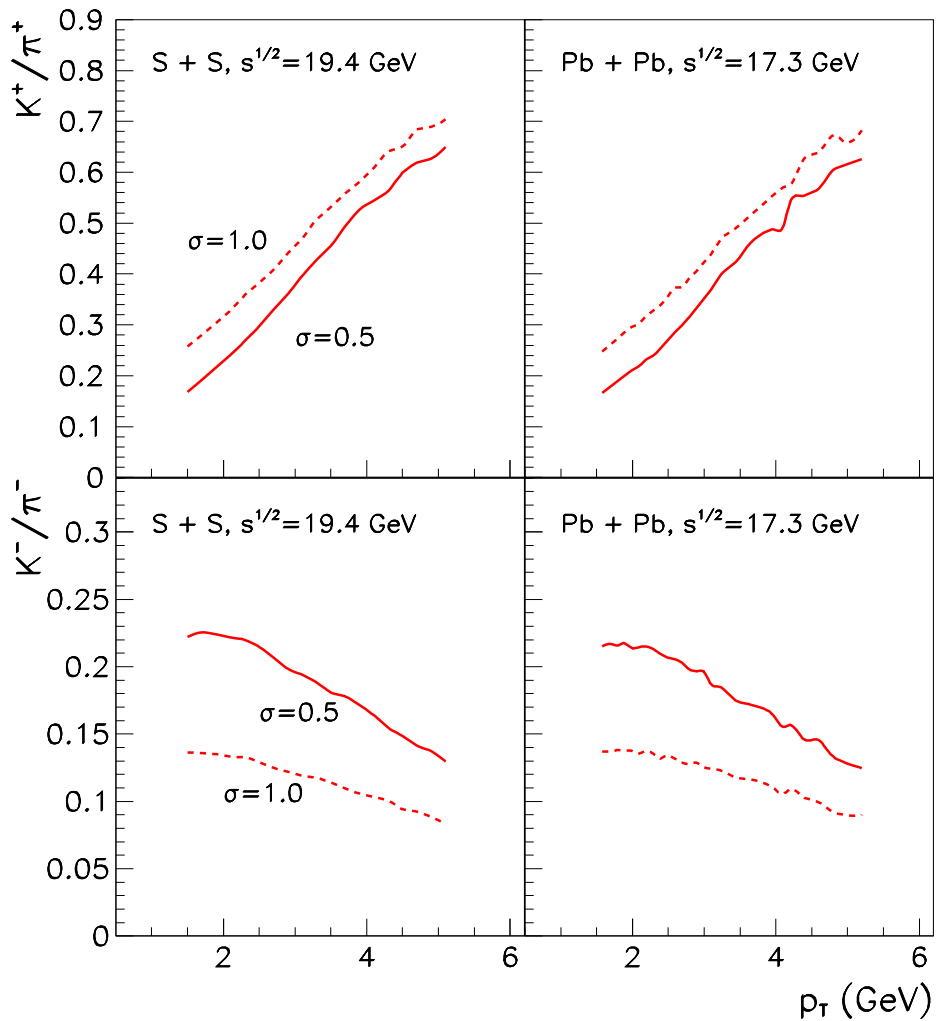}}
\vskip -0.05in
\caption[]{
 \label{figure11a}
Predicted $K^+/\pi^+$ (top) and $K^-/\pi^-$ (bottom) ratios 
for WA80 ($S + S$ at $\sqrt{s} = 19.4$ GeV,
left panels), and WA98 ($Pb + Pb$ at $\sqrt{s} = 17.3$ GeV, right 
panels) experimental conditions. The parameters of the calculation are fixed
as discussed in the text. Full lines 
correspond to $\sigma=0.5$, dashed lines display results with $\sigma=1$.
}
\end{figure}

We are not aware of hard ($p_T > 2$ GeV) kaon 
data from $AA$ collisions at SPS energies.
However, it is natural to extend the present calculations to kaon
production under the conditions of the WA80 and WA98 
experiments. Predicted $K^+/\pi^+$ (top) and $K^-/\pi^-$ (bottom) ratios 
for $S + S$ collisions at $\sqrt{s} = 19.4$ GeV
(left panels), and $Pb + Pb$ collisions at $\sqrt{s} = 17.3$~GeV (right
panels), as functions of $p_T$ are displayed in Fig. \ref{figure11a}.

The appearance of these results is similar to that of 
the calculated results in the $pA$ case, except that the curves 
are less steep for $AA$ collisions. Any possible peak at 
transverse momenta below $p_T=2$ GeV in the 
$K^-/\pi^-$ ratios would be unaccessible for our calculation. 
Information on these ratios can serve as a test of the present model at 
SPS energies. 

\section{Extension to RHIC energies} 
\label{sec_rhic}

In previous sections, we have developed a description of hard pion and 
kaon production based on the pQCD-improved parton model, incorporating
a phenomenological treatment of the transverse momentum degree of freedom 
of the partons. We have seen that a satisfactory agreement with $pp$ 
data can be achieved up to $\sqrt{s} \approx$ 60 GeV, allowing the 
width of the transverse-momentum distribution, $\k2av$, to depend 
on energy. The available significant 
body of information on hard pion production 
in the energy range 20 GeV $\lesssim \sqrt{s} \lesssim$ 60~GeV strongly
constrains the value of $\k2av$ at lower energies, but the uncertainty 
increases with $\sqrt{s}$. Modeling the enhancement of the
transverse-momentum width associated with additional collisions 
in the nuclear medium, we were able to reasonably reproduce 
hard pion-production data from $pA$ and $AA$ reactions in the same 
energy interval. We then illustrated the use of the model for kaons.

\subsection{Hadron production in $p\bar{p}$ collisions at energies 
$\sqrt{s}\gtrsim$ 100 GeV}
\label{sec_RHICpp}

In accordance with the main motivation for this study in terms of 
emerging data and future experiments at RHIC, in this Section the 
parameterization of the $pp$ data will be extended to the RHIC 
energy region. Unfortunately, we are not aware of useful {\it identified} 
hard meson production data  from $pp$ or $p\bar{p}$ reactions
at energies above $\sqrt{s} \approx$ 60 GeV; 
we have total hadron production data in the form of $(h^+ + h^-)/2$ from 
CERN UA1\cite{UA1arn,UA1alb,UA1boc} and the Tevatron CDF\cite{CDF}. Since the production 
of hadrons other than kaons and pions (in particular protons and antiprotons)
increases with energy, the accuracy of the extraction of $\k2av$ decreases
with increasing energy. Furthermore, we have to face a shortage of   
data in the energy interval 60 GeV $\lesssim \sqrt{s} \lesssim$ 500~GeV.  

\begin{figure}
\vskip -0.2in
\epsfxsize=5.5in
\epsfysize=6in
\centerline{\epsffile{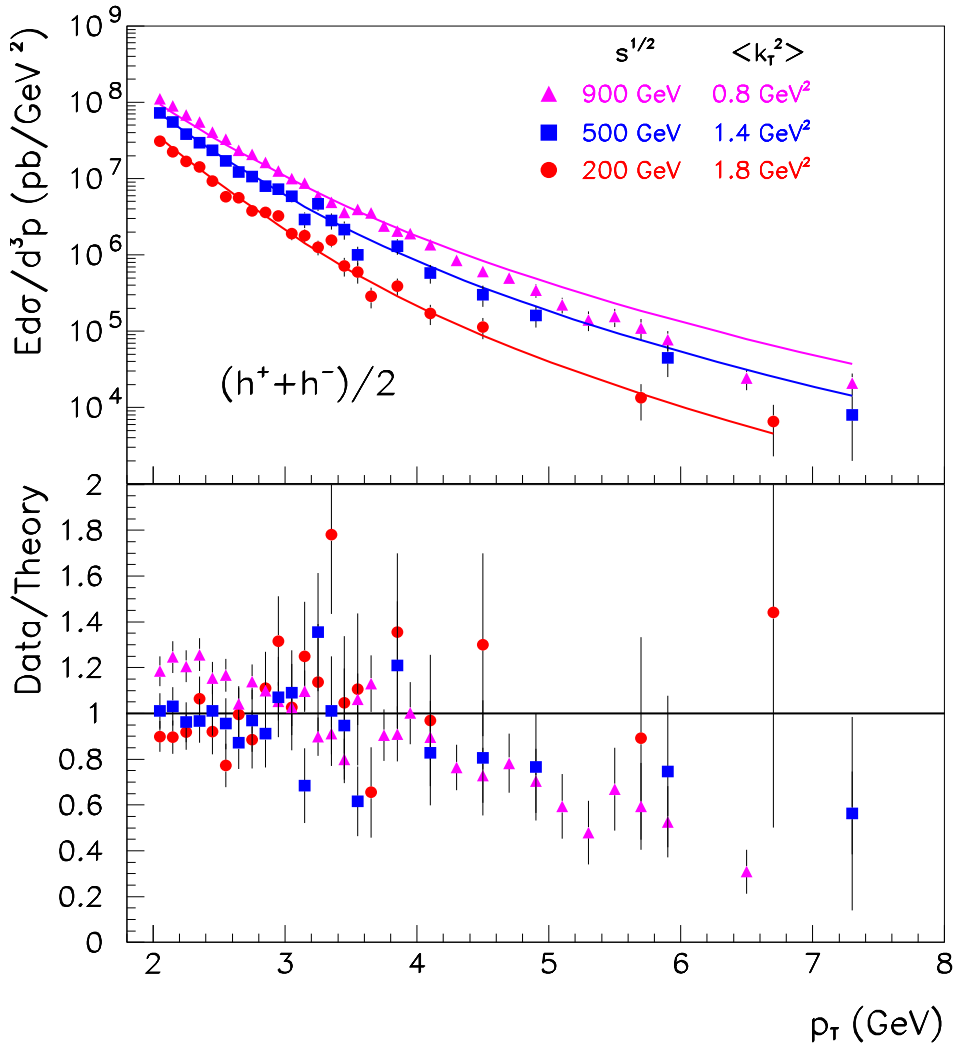}}
%\vskip -0.05in
\caption[]{
 \label{figure12}
Top panel: spectra of charged hadrons $(h^+ + h^-)/2$ from $p\bar{p}$ 
collisions at $\sqrt{s}=$200, 500, and 900 GeV as calculated 
with the indicated values of $\k2av$. Bottom panel: data/theory 
ratios. The data are from the UA1 experiment\cite{UA1alb}.
}
\end{figure}

In Fig. \ref{figure12} we show the quality of the fit we can achieve
with the energy-dependent $\k2av$ to the $(h^+ + h^-)/2$ 
data from $p\bar{p}$ collisions at $\sqrt{s}=$200, 500, and 900 GeV\cite{UA1alb}.
The top panel contains the spectra, the bottom panel gives the 
data/theory ratios as a function of $p_T$ for the values of $\k2av$
indicated in the top panel.  

\begin{figure}
%\vskip -0.4in
\epsfxsize=5.5in
\epsfysize=6in
\centerline{\epsffile{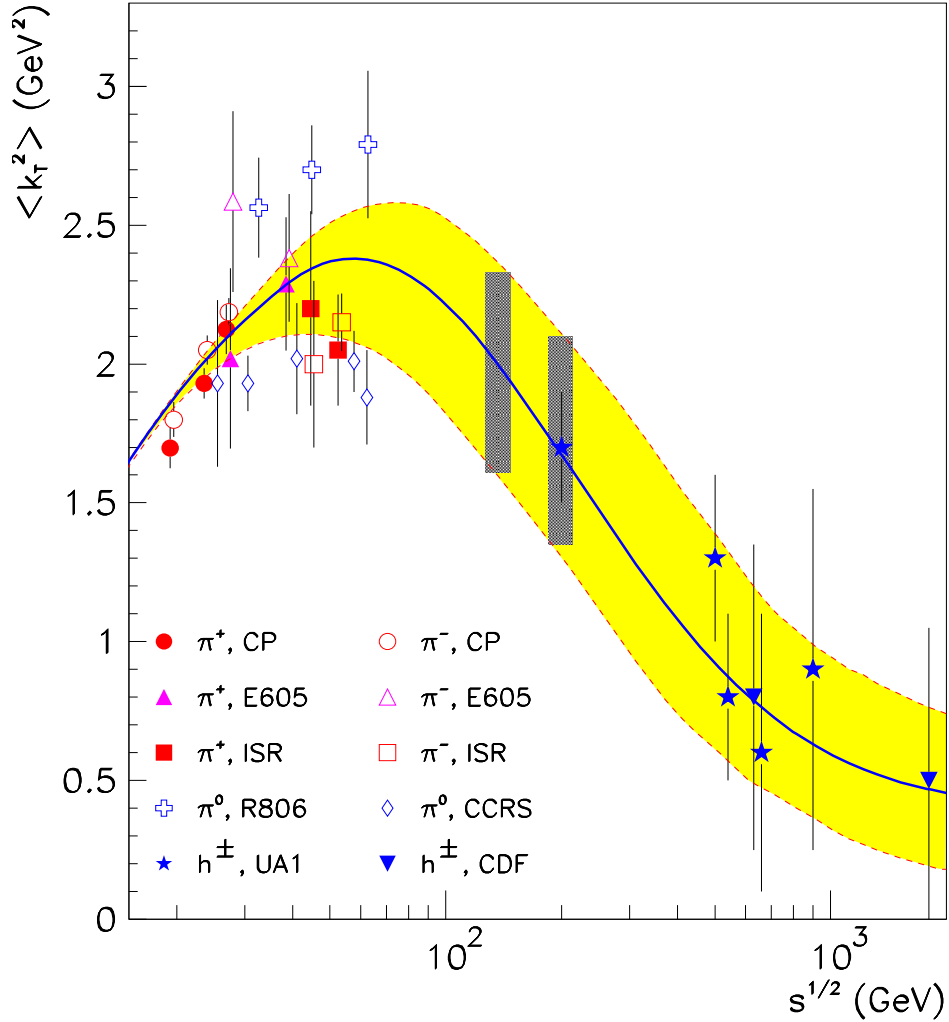}}
\vskip -0.05in
\caption[]{
 \label{figure13}
The best fit values of $\k2av$ in $p p \rightarrow \pi X$ 
\cite{antreasyan79,R806,CCRS,E605pi,E605pib,ISR}
%,E704pi,NA24pi,R110pi,WA70
and $p \bar{p} \rightarrow h^\pm X$ reactions at
$s^{1/2}=$200, 500, 540, 630, and 900 GeV \cite{UA1arn,UA1alb,UA1boc}
and $s^{1/2}=630, 1800$ GeV \cite{CDF}. 
Where large error bars would overlap at the same energy,
one of the points has been shifted slightly for better 
visibility. The band is drawn to guide the eye.
}
\end{figure}

Fig. \ref{figure13} summarizes the extracted values of the transverse 
momentum width from the $(h^+ + h^-)/2$ data of the
higher-energy experiments\cite{UA1arn,UA1alb,UA1boc,CDF}, together with the
points at lower energies extracted from hard {\it pion} production data
(shown earlier in Fig. \ref{figure3}). The solid curve is drawn to 
guide the eye, and is an attempt to represent the energy dependence
of the extracted $\k2av$. The dashed lines delineate
an estimated band of increasing uncertainty with 
increasing $\sqrt{s}$ around this curve.
We will use this band to interpolate 
between the low-energy and high-energy parts of the Figure,
reading off the $\k2av_{pp}$ values to be applied at RHIC. 
The darker bars represent an estimate at around $\sqrt{s}=$ 130 and 200 
GeV, respectively.

It can be seen that $\k2av$ reaches a maximum at around 
$\sqrt{s}=$ 60-100 GeV, with a value of $\approx 2.4$ GeV$^2$.
Beyond this energy, the transverse-momentum width necessary
to reproduce hadron production data from $pp$ or $p\bar{p}$
reactions at the LO level of the pQCD calculation starts to
decrease.
The interpolating curve is drawn so that it
converges to a small constant value of $\k2av$
as $\sqrt{s} \longrightarrow \infty$. 
The high energy points are based on charged pion and kaon production
only, and should be looked upon as upper limits.
In the RHIC energy region, we extract $\k2av=2.0 \pm 0.4$ at 
$\sqrt{s} \approx $ 130 GeV
and $\k2av=1.7 \pm 0.4$ at $\sqrt{s} \approx $ 200 GeV for the value of the 
transverse-momentum width of partons without nuclear enhancement.

For the $pA$ and $AA$ calculations we fixed $\nu_m=4$ and $C=0.4$ GeV$^2$, 
in lack of a less arbitrary prescription.
It needs to be kept in mind, however, that these quantities and the 
effectivity function $h_{pA}(b)$ of Eq. \ref{ktbroadpA} may well be 
energy dependent. We carried out computations for pion and kaon 
production at the RHIC energies of the 2000 run and of the 2001 run
of $\sqrt{s}=$130 GeV and 200 GeV, respectively.

\subsection{Hadron production in $pA$ collisions at RHIC energies}
\label{sec_RHICpA}

\begin{figure}
\vskip -0.4in
\epsfxsize=5.in
\epsfysize=5.5in
\centerline{\epsffile{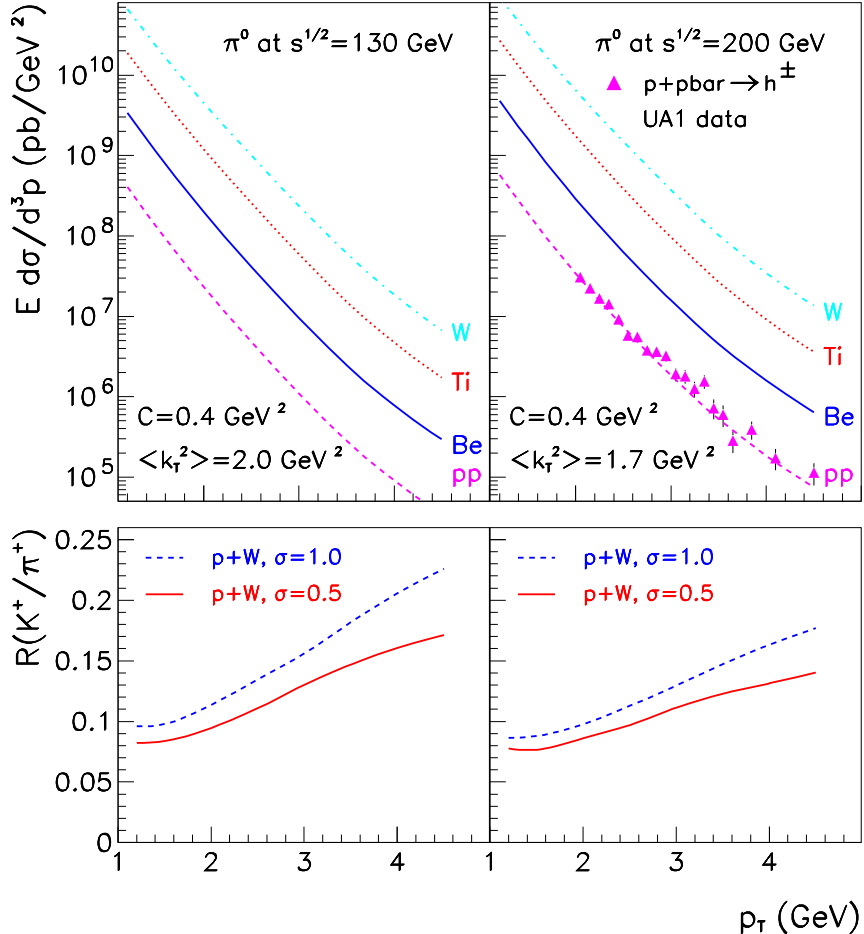}}
\vskip -0.05in
\caption[]{
 \label{figure13a}
Top panels: predicted invariant cross sections for $\pi^0$ production 
from $p + Be, Ti, W$ collisions at $\sqrt{s}=$ 130 GeV (left) and 
200 GeV (right). Calculated $pp$ results and the UA1 $h^\pm$ data
from $p\bar{p}$ collisions\cite{UA1alb} are also included.
Bottom panels: predicted $K^+/\pi^+$ ratios for $p + W$
collisions with $\sigma =$ 0.5(solid) and $\sigma =$ 1.0 (dashed). 
}
\end{figure}

We are now prepared to make certain predictions concerning 
the planned $pA$ program at RHIC\cite{pARHIC}. Since the precise
energies and targets to be used are not presently available, 
we use the c.m. energies of $\sqrt{s}=$ 130 and 200 GeV, 
respectively, and the targets familiar from Section \ref{sec_nu}, as an
illustration. For the width of the transverse momentum distribution of
partons in the proton, we take the values $\k2av=$ 2.0 GeV$^2$ and 1.7 GeV$^2$ at
$\sqrt{s}=$130 and 200 GeV, respectively, representing the centers of
the dark bars in Fig. \ref{figure13}. The parameters 
$\nu_m = 4$ and $C = 0.4$~GeV$^2$ are fixed. In the top panels 
of Fig. \ref{figure13a} we show calculated invariant $\pi^0$
production cross sections at these energies from collisions of protons
with $Be$ (solid line), $Ti$ (dotted), and $W$ (dot-dash) targets.
We also include the results of $pp$ calculations as a reference, and 
the UA1 $p\bar{p}$ data at $\sqrt{s}=$ 200 GeV for $h^\pm$ production
for comparison\cite{UA1alb}. These data were used earlier      
in Fig. \ref{figure12}. Since kaon production typically does 
not amount to more than 20\% of pion production and the use of antiprotons 
is not expected to make a significant difference at these energies,
the $p+p \longrightarrow \pi^0 + X$ calculation appears to fit these 
data on the logarithmic scale almost as well as the earlier $h^\pm$
calculation (see Fig. \ref{figure12}). In the bottom panels 
of Fig. \ref{figure13a} $K^+/\pi^+$ ratios are displayed for 
the $p + W$ reactions, at the values of $\sigma=0.5$ (solid line)
and $\sigma=1.0$ (dashed) for the parameter controlling sea-quark
fragmentation. These ratios are very similar for the other targets.

\begin{figure}
\vskip -0.4in
\epsfxsize=5.5in
\epsfysize=6in
\centerline{\epsffile{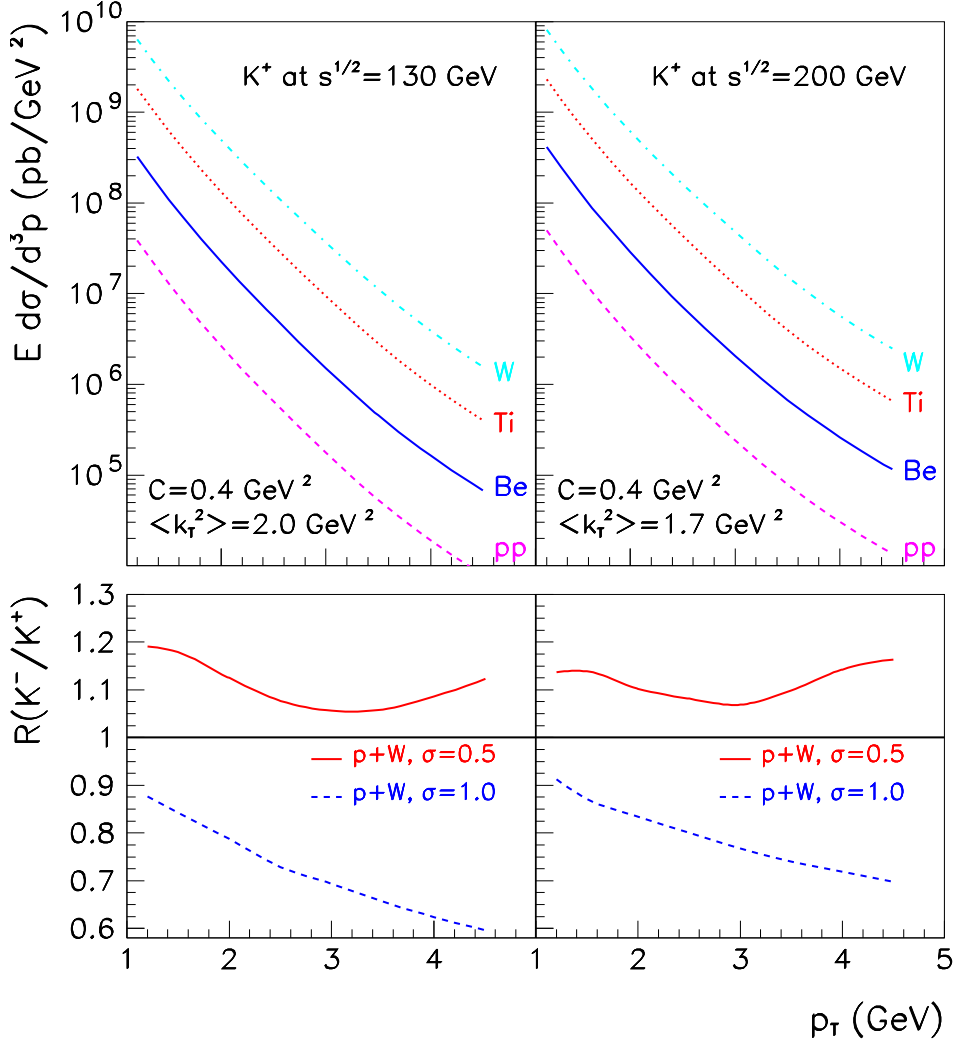}}
\vskip -0.05in
\caption[]{
 \label{figure13b}
Top panels: predicted invariant cross sections for $K^+$ production 
from $p + Be, Ti, W$ collisions at $\sqrt{s}=$ 130 GeV (left) and 
200 GeV (right). Calculated $pp$ results are also included.
Bottom panels: predicted $K^-/K^+$ ratios for $p + W$
collisions with $\sigma =$ 0.5(solid) and $\sigma =$ 1.0 (dashed). 
}
\end{figure}

Figure \ref{figure13b} displays similar predictions for $K^+$ 
production from $pA$ collisions. In the bottom panels we now 
show $K^-/K^+$ ratios in $p + W$ collisions for the two 
values of the parameter $\sigma$ used earlier. It can be seen 
from the bottom panels 
that the value of $\sigma$ plays a more important role in kaon 
than in pion production also at these higher energies.

\subsection{Hadron production in $Au+Au$ collisions at RHIC}
\label{sec_RHICAA}

\begin{figure}
%\vskip -0.4in
\epsfxsize=5.5in
\epsfysize=6in
\centerline{\epsffile{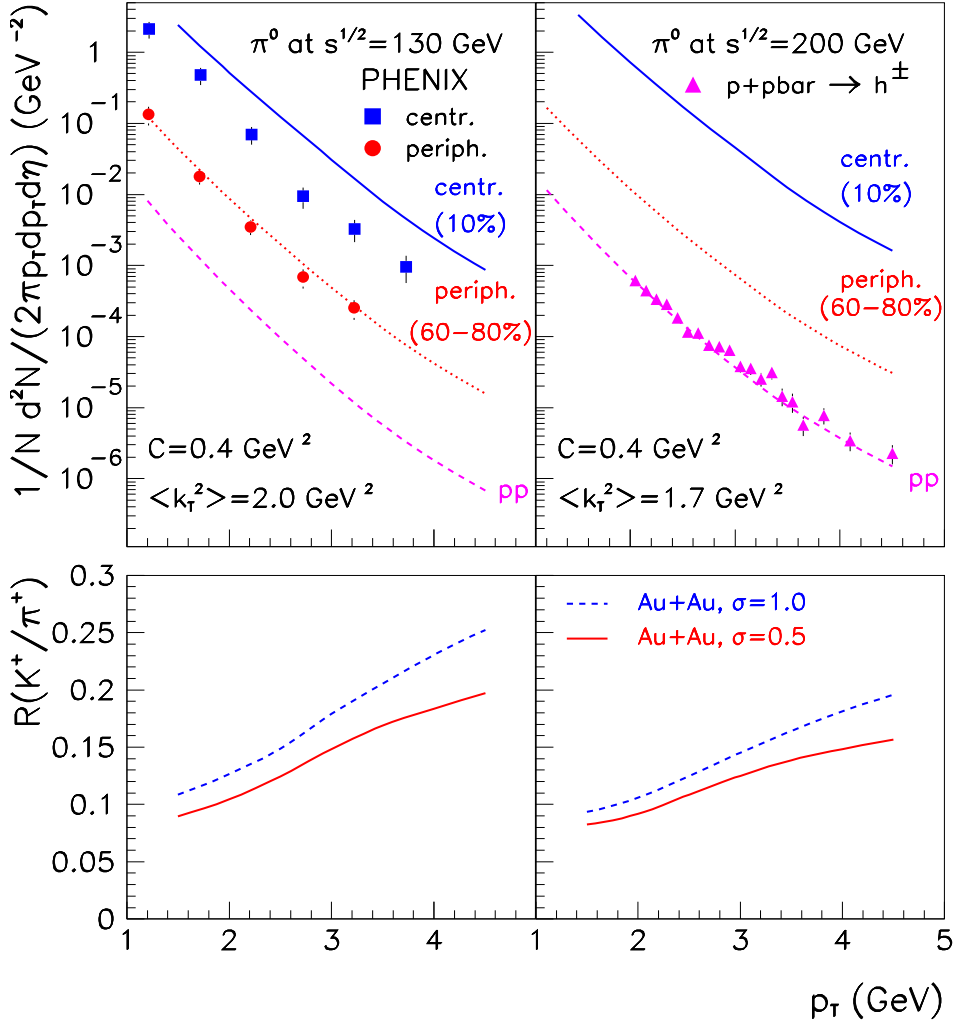}}
\vskip -0.05in
\caption[]{
 \label{figure14}
Top left: calculated central 10\% (full line) and peripheral 
60-80\% (dotted) hard $\pi^0$
spectra from $Au+Au$ collisions at $\sqrt{s}=$130 GeV, compared to preliminary
PHENIX data\cite{david00}. We also show the result of our $pp$ calculation
(dashed). Top right: prediction for central (full line) and peripheral (dotted) 
hard $\pi^0$ spectra from $Au+Au$ collisions at $\sqrt{s}=$200 GeV. The 
$pp$ results are included and compared to the UA1 $p\bar{p}$ data at the 
same energy\cite{UA1alb}. Bottom: predicted $K^+/\pi^+$ ratios at the same
energies.   
}
\end{figure}

In the area of $AA$ collisions, calculated high-$p_T$
$\pi^0$ production in $Au+Au$ collisions is compared 
to preliminary PHENIX data at $\sqrt{s}=$130 GeV\cite{david00}
in the top left panel of Fig. \ref{figure14};
predictions at 200 GeV are displayed on  
the right of Fig. \ref{figure14}. 
We use the values $\k2av=$ 2.0 GeV$^2$ and 1.7 GeV$^2$ at
$\sqrt{s}=$130 and 200 GeV, respectively, representing the centers of
the dark bars in Fig. \ref{figure13}. The parameters 
$\nu_m = 4$ and $C = 0.4$~GeV$^2$ are fixed.    
Central and peripheral collisions are defined by the top 10\%
and the 60-80\% bin of the total cross section, corresponding 
to the experimental selection. We also show the results of 
the pQCD calculations of hard pion production from $pp$ collisions 
for reference. The $p\bar{p}$ data on total 
charged hadron production at 200 GeV\cite{UA1alb} are also included 
in the top right panel of the Figure (symbols).
In the bottom panels the predicted $K^+/\pi^+$ ratios 
associated with central collisions are displayed.
The corresponding peripheral ratios 
are almost identical to the central ratios shown.

It can be seen in the top left panel of Fig. \ref{figure14}
that, while our pQCD calculation augmented by nuclear effects
approximately reproduces the data 
on peripheral $Au + Au$ collisions at $\sqrt{s}=$130 GeV,
the central data are overestimated by a factor of $3-5$.
This indicates that an additional mechanism is at work in dense
nuclear matter, which would decrease the calculated cross sections
at a given $p_T$. Since this effect can also be looked upon as a 
shift of the spectra to lower $p_T$, we speculate that the
phenomenon of jet quenching\cite{plumer,baier98,gyulassy00}
is responsible for the discrepancy in central collisions.
A study of jet quenching
in the present framework is in progress\cite{levai01}. 

\begin{figure}
%\vskip -0.4in
\epsfxsize=5.5in
\epsfysize=6in
\centerline{\epsffile{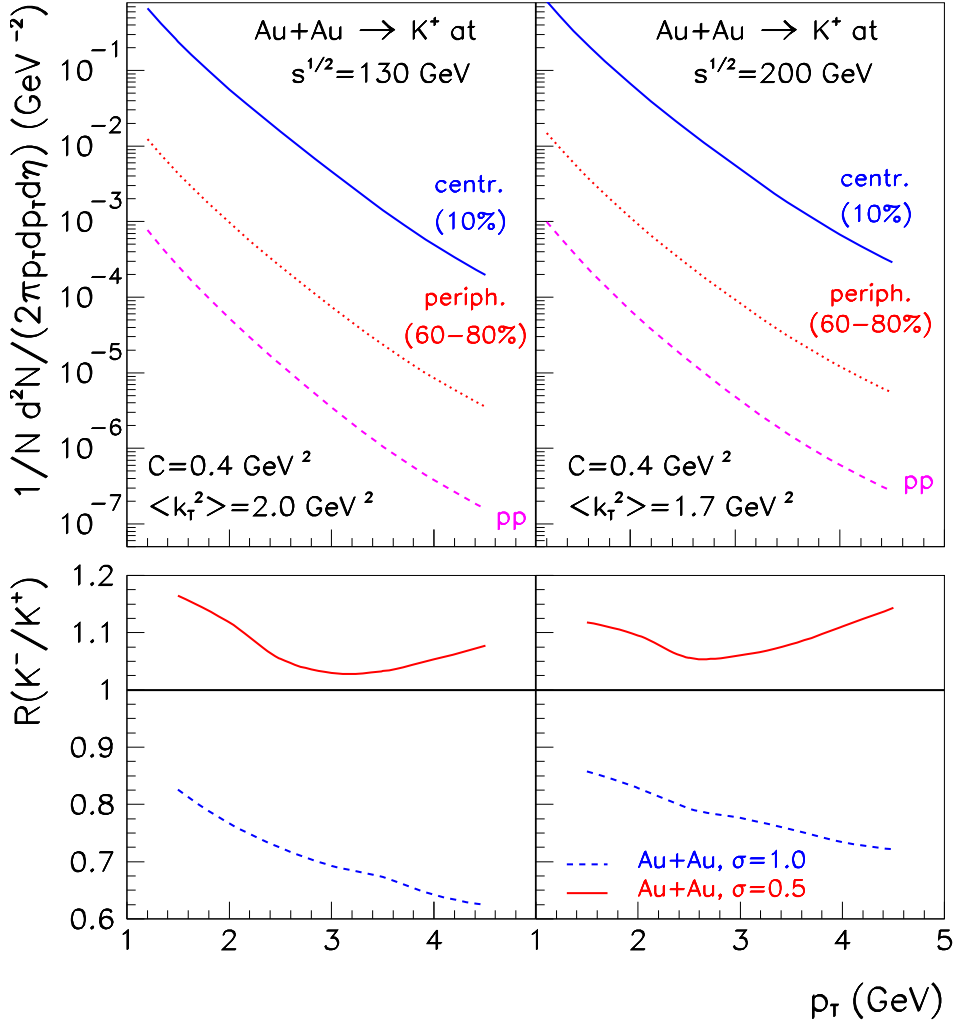}}
\vskip -0.05in
\caption[]{
 \label{figure15}
Predicted  K$^+$ spectra and for $K^-/K^+$
ratios for $Au+Au$ reactions at $\sqrt{s}=$ 130 GeV (left panels)
and 200 GeV (right panels). 
}
\end{figure}

The predictions in Sections \ref{sec_RHICpA} and 
\ref{sec_RHICAA} above are intended to illustrate 
the capabilities of the description we developed. Similar results can be obtained easily
for the precise conditions of future RHIC measurements. We find a systematic study 
of the energy dependence of hard pion and kaon production in $pp$ collisions 
particularly important. Similarly, a study of the energy and target
dependence of pion and kaon production in $pA$ collisions 
will help pin down the energy dependence of the parameters
that characterize the enhancement of the transverse momentum width in the medium,
which have been taken energy independent for the time being.

%\newpage
\section{Summary and conclusion}
\label{sec_sum}

The commissioning of RHIC opened an exciting new era in nuclear 
collision physics, with the study of excited strongly-interacting matter 
becoming a reality. It is recognized, however, that a thorough 
understanding of ultrarelativistic $AA$ collisions presupposes the 
accurate description of $pp$ and $pA$ collisions in the same 
framework. In the present paper we attempted to follow the evolution 
of hard pion and kaon production in relativistic collisions from $pp$
to $pA$ to $AA$ reactions.

The pQCD-improved parton model suggested itself as a natural tool for
our study. Of course, pQCD itself is evolving from leading-order calculations
through the increasing complexity of NLO and NNLO for selected processes.
As our major focus in this paper was on the additional physics brought in by $pA$ and
$AA$ collisions, it was decided to use LO pQCD throughout. This was
supported by evidence that higher order pQCD does not eliminate the 
need for the transverse momentum distribution of partons for a satisfactory
fit of pion and kaon production data in $pp$ collisions 
in the $2 \leq p_T \leq 6$ GeV window (though the    
numerical values become smaller, as expected). We used abundant pion production 
data from $pp$ collisions at c.m. energies $\sqrt{s} \lesssim $ 60 GeV   
to extract the width of the transverse momentum distribution of 
partons in the nucleon. The phenomenological value of the 
description was then tested on kaon production at the same energies.

For the treatment of nuclear systems, we developed a model based 
on the enhancement of the width of the transverse momentum distribution of 
partons in the nuclear medium. An additional parameter 
was fitted to describe the Cronin effect at these energies. Shadowing
and the isospin asymmetry of heavy nuclei were  taken into account.
We tested the model on charged pion and kaon production. In $AA$ 
collisions at SPS energies we found an indication of a possible 
need for an additional mechanism to decrease the calculated cross
section of neutral pion production in the collision of heavy nuclei.
We speculated that jet quenching may provide that mechanism,
beginning to appear already at SPS energies.

Using $p\bar{p}$ data at higher energies, the domain of the 
extracted transverse momentum width was extended to cover 
planned and present RHIC energies. Keeping in mind the rather 
large uncertainties, predictions were made for $pA$ and $AA$ 
collisions at RHIC. The over-prediction of the preliminary PHENIX data 
on central $\pi^0$ production at $\sqrt{s}=$ 130 GeV reinforced our 
tentative conclusion that additional physics needs to be 
incorporated in the model to decrease the calculated cross 
sections when dense nuclear matter is present. Since the 
disagreement with the central data is much stronger at these 
energies than at SPS, and jet quenching is expected to increase 
with energy, jet quenching is a likely candidate for this physics.

Recognizing that the treatment of jet energy loss would require further 
assumptions and modeling, we leave it for later study. A future treatment of 
jet quenching in the same framework promises to be of interest
as a measure of the gluon density of the medium responsible for the  
energy loss. We are also interested in a similar study at the NLO
level of pQCD.

In conclusion, it appears that a description of $pA$ and $AA$ 
collisions based on the pQCD-improved parton model augmented by 
the transverse momentum distribution of partons and by the nuclear 
effects of multiscattering and shadowing works reasonably well 
as a background calculation for hard pion and kaon production
at sufficiently high energies and has useful predictive power
for the interpretation of present and future RHIC data. A full treatment of 
nuclear effects will need to take jet energy loss into account.
We see our model as a flexible tool to aid in the understanding of 
the properties of extended strongly-interacting matter 
created at RHIC, LHC, and potential future nuclear colliders.

\vspace*{4 mm}

\section{Acknowledgements}
\label{sec_ack}

We thank B. Cole, G. David, M. Gyulassy, M. Tannenbaum, G. Odyniec, T. Peitzmann,
M. Corcoran, X-N.~Wang, and C.Y. Wong for useful comments and stimulating discussions. 
This work was supported in part by  U.S. DOE grant DE-FG02-86ER40251, NSF grant 
INT-0000211, FKFP220/2000 and Hungarian grants OTKA-T032796 and OTKA-T034842. 
Supercomputer time provided by BCPL in Bergen, Norway is gratefully acknowledged. 

\vspace*{4 mm}

\section{Appendix: Fragmentation functions for charge-separated pions and kaons}
\label{sec_app}

The parameterization of fragmentation functions (FFs) for pions and kaons is 
usually given in terms of neutral linear combinations of the charged 
mesons\cite{BKKII,KKP}, denoted by e.g. $D^{\pi^{\pm}}_{u}$, $D^{\pi^{\pm}}_{d}$, and
$D^{\pi^{\pm}}_{s}$ for pions. Since the present paper focuses attention on separately 
measured $\pi^+$, $\pi^-$, $K^+$, and $K^-$ spectra and their ratios,
we need separate FFs for the charged pions and kaons. In this Appendix 
we describe our approximation used for the charge-separated pion and kaon FFs.

To lowest order, one may assume that each meson species will only fragment
from those quarks that appear in it as valence quarks, or from gluons. In this
approximation, there is e.g. no contribution to the fragmentation of pions
from strange quarks, or to the fragmentation of kaons from $d$ or $\bar{d}$
quarks. However, these flavors can appear as sea quarks in the respective mesons,
and we want to take into account the possibility that pions or kaons fragment from 
quarks that appear in them as sea contributions. This implies a separation of
the FFs into valence and sea parts, along the lines of Ref. \cite{BKKI}. 
For consistency, 
the sea quarks corresponding to the valence flavors in the given meson need also 
be considered. 

Based on the above ideas, we split the fragmentation functions for pions as
\begin{eqnarray}
D^{\pi^{\pm}}_{u}  =  & D^{\pi^{\pm}}_{u,val}  + & D^{\pi^{\pm}}_{u,sea} \\
D^{\pi^{\pm}}_{d}  =  & D^{\pi^{\pm}}_{d,val}  + & D^{\pi^{\pm}}_{d,sea} \\
D^{\pi^{\pm}}_{s}  =  &                          & D^{\pi^{\pm}}_{s,sea} \,\, .
\end{eqnarray}

The sea contribution from the fragmentation of a $u$ or $d$ quark can be 
identified in this case with the FF of the $s$ quark, which appears in 
the pion only as a sea quark,
\begin{equation}
D^{\pi^{\pm}}_{u,sea} = D^{\pi^{\pm}}_{d,sea} =
D^{\pi^{\pm}}_{s,sea} = D^{\pi^{\pm}}_{s}   \,\, .
\end{equation}           
Furthermore, it is natural to make the usual ansatz connecting
the charge-averaged FFs of quarks and antiquarks\cite{BKKI}:
\begin{equation}
D^{\pi^{\pm}}_{\bar{u}} = D^{\pi^{\pm}}_{u}  \,\, ,
\end{equation}
and similarly for $d$ and $s$ quarks and antiquarks.

The valence contributions to the $\pi^{\pm}$ FFs can be 
obtained from the parameterization available in Ref.~\cite{KKP} as 
\begin{eqnarray}
D^{\pi^{\pm}}_{u,val} & =  & D^{\pi^{\pm}}_{u} - D^{\pi^{\pm}}_{s}  \\
D^{\pi^{\pm}}_{d,val} & =  & D^{\pi^{\pm}}_{d} - D^{\pi^{\pm}}_{s} \,\, .
\end{eqnarray}
Since $u$ and $\bar{d}$ occur in the positive pion as valence quarks, we have 
\begin{eqnarray}
D^{\pi^{+}}_{u,val} & = & D^{\pi^{\pm}}_{u,val} \\
D^{\pi^{+}}_{\bar{d},val} & = & D^{\pi^{\pm}}_{\bar{d},val}
\,\, ,
\end{eqnarray}
and similarly for $\pi^-$.  

The remaining task is to divide the contribution of $u$ quark fragmentation to
pions between $\pi^+$ and $\pi^-$ when they appear 
in the created pion as sea quarks. We propose 
\begin{eqnarray}
D^{\pi^{+}}_{u,sea} = & \sigma & D^{\pi^{\pm}}_{u,sea}  \\
D^{\pi^{-}}_{u,sea} = & (1-\sigma) & D^{\pi^{\pm}}_{u,sea} \\
D^{\pi^{+}}_{\bar{u},sea} = & (1-\sigma) & D^{\pi^{\pm}}_{\bar{u},sea}  \\
D^{\pi^{-}}_{\bar{u},sea} = & \sigma & D^{\pi^{\pm}}_{\bar{u},sea}
\end{eqnarray} 
where $0 \leq \sigma \leq 1$ is a free parameter. It is natural to expect
$\sigma=0.5$ from symmetry arguments. We find that $\sigma=0.5$ indeed 
provides a satisfactory description of the available pion data,
and use this as the default value in the present paper.

For charged kaons a very similar analysis, with the $s$ quark replacing the
$d$ quark as a valence contribution and the $d$ quark playing the 
role of the contribution that can only arise from the sea, defines the
fragmentation functions in an analogous manner.
 
%\vfill\eject\newpage


\begin{thebibliography}{100}

\bibitem{QM01}
Proceedings of {\it Quark Matter 2001}, Nucl. Phys. A {\bf xxx}, 1 (2001);
http://www.rhic.bnl.gov/qm2001.

\bibitem{owens87}
J.F. Owens, Rev. Mod. Phys. {\bf 59}, 465 (1987).
%% CITATION = RMPHA,59,465;%%

\bibitem{FF95}
R.D. Field, {\it Applications of Perturbative QCD}, Addison-Wesley, 1995.

\bibitem{satz01}
H. Satz, Nucl. Phys. Proc. Suppl. {\bf 94}, 204 (2001);
R. Vogt, hep-ph/0107045 (2001).
%% CITATION = HEP-PH 0107045;%%

\bibitem{leit00}
M.J. Leitch {\it et al.} (E866/NuSea), Phys. Rev. Lett. {\bf 84}, 3256 (2000).
%% CITATION = PRLTA,84,32565;%%

\bibitem{cronin75}
J.W. Cronin, H.J. Frisch, M.J. Shochet, 
J.P. Boymond, P.A. Piroue, and R.L. Sumner (CP),
Phys. Rev. D {\bf 11}, 3105 (1975).
%% CITATION = PHRVA,D11,3105;%%

\bibitem{antreasyan79}
D. Antreasyan, J.W. Cronin, H.J. Frisch, M.J. Shochet, L.Kluberg, P.A. Piroue, and
R.L. Sumner (CP), Phys. Rev. D {\bf 19}, 764 (1979).
%% CITATION = PHRVA,D19,764;%%

\bibitem{R806}
C. Kourkoumelis {\it et al.} (R806), Z. Phys. C {\bf 5}, 95 (1980).
%% CITATION = ZEPYA,C5,95;%%

\bibitem{CCRS}
F.W. B{\"u}sser {\it et al.} (CCRS), Nucl. Phys. B {\bf 106}, 1 (1976).
%% CITATION = NUPHA,B106,1;%%

\bibitem{E605pi}
D.E. Jaffe {\it et al.} (E605), Phys. Rev. D {\bf 40,} 2777 (1989).
%% CITATION = PHRVA,D40,2777;%%

\bibitem{E605pib}
P.B. Straub {\it et al.}, Phys. Rev. Lett. {\bf 68}, 452 (1992).
%% CITATION = PRLTA,68,452;%%

\bibitem{ISR}
B. Alper {\it et al.} (ISR), Nucl. Phys. B {\bf 100}, 237 (1975).
%% CITATION = NUPHA,B100,237;%%

\bibitem{UA1arn}
G. Arnison {\it et al.} (UA1), Phys. Lett. B {\bf 118}, 167 (1983).
%% CITATION = PHLTA,B118,167;%%

\bibitem{UA1alb}
C. Albajar {\it et al.} (UA1), Nucl. Phys. B {\bf 335}, 261 (1990).
%% CITATION = NUPHA,B335,261;%%

\bibitem{UA1boc}
C. Bocquet {\it et al.} (UA1), Phys. Lett. B {\bf 366}, 434 (1996). 
%% CITATION = PHLTA,B366,434;%%

\bibitem{CDF}
F. Abe {\it et al.} (CDF), Phys. Rev. Lett. {\bf 61}, 1819 (1988).
%% CITATION = PRLTA,61,1819;%%

\bibitem{brown96}
C.N. Brown, {\it et al.}, Phys. Rev. C {\bf 54}, 3195 (1996).
%% CITATION = PHRVA,C54,3195;%%

\bibitem{ApaBep} L. Apanasevich {\it et al.} (E706),
        Phys. Rev. Lett. {\bf 81}, 2642 (1998);
        Phys. Rev. D {\bf 63}, 014009 (2000).
%% CITATION = PRLTA,61,1819;%%
%% CITATION = PHRVA,D63,014009;%%

\bibitem{plumer}
M. Gyulassy and M. Pl\"umer, Phys. Lett. B {\bf 243}, 432 (1990);
X.-N. Wang and M. Gyulassy, Phys. Rev. Lett. {\bf 68}, 1480 (1992)
%% CITATION = PHLTA,B243,432;%%
%% CITATION = PRLTA,68,1480;%%

\bibitem{baier98}
R. Baier, Yu.L. Dokshitzer, A.H. Mueller, and D. Schiff, Nucl. Phys. B
{\bf 531}, 403 (1998);
R. Baier, D. Schiff, B.G. Zakharov, Ann. Rev. Nucl. Part. Sci.
{\bf 50}, 37 (2000).
%% CITATION = ARNUA,50,37;%%
%% CITATION = NUPHA,B531,403;%%

\bibitem{gyulassy00}
M. Gyulassy, P. L{\'e}vai, and I. Vitev, Phys. Rev. Lett. {\bf 85}, 
5535 (2000); Nucl. Phys. B {\bf 594}, 371 (2001).
%% CITATION = PRLTA,85,5535;%%
%% CITATION = NUPHA,B594,371;%%

\bibitem{wang01;a}
M. Gyulassy, I. Vitev, and X.-N. Wang, Phys. Rev. Lett. {\bf 86}, 2537 (2001);
X.-N. Wang and M. Gyulassy, Phys. Rev. Lett. {\bf 86}, 3496 (2001).
%% CITATION = PRLTA,86,2537;%%
%% CITATION = PRLTA,86,3456%%

\bibitem{plf00}
G. Papp, P. L{\'e}vai, and G. Fai, Phys. Rev. C {\bf 61}, 021902(R) (2000).
%% CITATION = PHRVA,C61,021902;%%

\bibitem{cole}
B. Cole {\it et. al.} (E910), Nucl. Phys. A {\bf 661}, 366 (1999).
%% CITATION = NUPHA,A661,366;%%

\bibitem{Chem}
I. Chemakin {\it et al.} (E910), Phys. Rev. Lett. {\bf 85}, 4868 (2000);
nucl-ex/0108007.
%% CITATION = PRLTA,85,4868;%%
%% CITATION = HEP-PH/0108007;%%

\bibitem{Cole01}
B. Cole, 
Proceedings of {\it Quark Matter 2001}, Nucl. Phys. A {\bf xxx}, 1 (2001);
http://www.rhic.bnl.gov/qm2001

\bibitem{Veres01}
G. Veres for NA49 Collaboration, poster at the Conference
{\it Quark Matter 2001}; http://www.rhic.bnl.gov/qm2001.

\bibitem{pARHIC}
S. Aronson {\it et al.},  White paper on pA physics at RHIC,
http://www.bnl.gov/rhic/townmeeting/agenda\_b.htm (2000).

\bibitem{wang01;b}
X.N. Wang, Phys. Rev. C {\bf 61}, 064910 (2001). 
%% CITATION = PHRVA,C61,064910;%%

\bibitem{KKP}
B.A. Kniehl, G. Kramer, and B. P{\"o}tter, Nucl. Phys. B {\bf 597}, 337 (2001);
hep-ph/0011155
%% CITATION = NUPHA,B597,337;%%
%% CITATION = HEP-PH 0011155;%%

\bibitem{bflpz01}
G.G. Barnaf\"oldi, G. Fai, P. L{\'e}vai, G. Papp, and Y. Zhang,
J. Phys. G {\bf 27}, 1767 (2001).

\bibitem{E609}
M.D. Corcoran {\it et al.} (E609), Phys. Lett. B {\bf 259}, 209 (1991).
%% CITATION = PHLTA,B259,209;%%

\bibitem{GRV92}
M. Gl{\"u}ck, E. Reya, and A. Vogt, Z. Phys. C {\bf 53}, 127 (1992).
%% CITATION = ZEPYA,C53,127;%% 

\bibitem{aurenche99}
P. Aurenche, M. Fontannaz, J.Ph. Guillet, B. Kniehl, E. Pilon, and M. Werlen,
Eur. Phys. J. C {\bf 9}, 107 (1999);
P. Aurenche, M. Fontannaz, J.Ph. Guillet, B. Kniehl, and M. Werlen,
Eur. Phys. J. C {\bf 13}, 347 (2001).
%% CITATION = EPHJA,C9,107;%%
%% CITATION = EPHJA,C13,347;%%

\bibitem{pzfbl02}
G. Papp, Y. Zhang, G. Fai, G.G. Barnaf\"oldi, and P. L\'evai, in preparation

\bibitem{stevenson81}
P.M. Stevenson, Phys. Rev. D {\bf 23}, 2916 (1981).
%% CITATION = PHRVA,D23,2916;%%

\bibitem{BKKI}
J. Binnewies, B.A. Kniehl, and G. Kramer, Z. Phys. C {\bf 65}, 471 (1995). 
%% CITATION = ZEPYA,C65,471;%% 

\bibitem{kimber00}
M.A. Kimber, A.D. Martin, and M.G. Ryskin, Eur. Phys. J. C {\bf 12}, 655 (2000);
hep-ph/0101348 (2001).
%% CITATION = EPHJA,C12,655;%%
%% CITATION = HEP-PH 0101348;%%

\bibitem{Wang9798}
X.N. Wang,
        Phys. Rep. {\bf 280}, 287 (1997);
        Phys. Rev. Lett. {\bf 81,} 2655 (1998);
        Phys. Rev. C {\bf 58}, 2321 (1998).
%% CITATION = PRPLC,280,287;%%
%% CITATION = PRLTA,81,2655;%%
%% CITATION = PHRVA,C58,2321;%%

\bibitem{Wong98}
C.Y. Wong and H. Wang,
        Phys. Rev. C {\bf 58,} 376 (1998).
%% CITATION = PHRVA,C58,376;%%

\bibitem{sivers76}
D. Sivers, S. Brodsky, and R. Blankenbecler,
        Phys. Rep. {\bf 23}, 1 (1976);
A.P. Contogouris, R. Gaskell and S. Papadopoulos,
        Phys. Rev. D {\bf 17}, 2314 (1978).
%% CITATION = PRPLC,280,287;%%
%% CITATION = PHRVA,D17,2314;%%

\bibitem{Guo96}
X. Guo and J. Qiu,
        Phys. Rev. D {\bf 53}, 6144 (1996).
%% CITATION = PHRVA,D53,6144%%

\bibitem{Lai98}
H.L. Lai and H.N. Li,
        Phys. Rev. D {\bf 58}, 114020 (1998).
%% CITATION = PHRVA,D58,114020%%

\bibitem{Hust} J. Huston {\it et al.},
        Phys. Rev. {\bf D51}, 6139 (1995). 
%% CITATION = PHRVA,D51,6139;%%

\bibitem{sridhar98}
K. Sridhar, A.D. Martin, and W.J. Stirling, Phys. Lett. B {\bf 438}, 211 (1998).
%% CITATION = PHLTA,B438,211;%%

\bibitem{zhuang01}
P. Zhuang and J. H{\"u}fner, nucl-th/0109037.
%% CITATION = NUCL-TH 0109037;%%

\bibitem{nikolaev99}
N.N. Nikolaev, hep-ph/9905562.
%% CITATION = HEP-PH 9905562;%%

\bibitem{VEGAS}
R. Kreckel, physics/9812011; G.P. Lepage, J. Comput. Phys. {\bf 27}, 192 (1978).
%% CITATION = PHYSICS 9812011;%%
%% CITATION = JCTPA,27,192;%%

\bibitem{kuhn76}
J.H. K{\"u}hn, Phys. Rev. D {\bf 13}, 2948 (1976).
%% CITATION = PHRVA,D13,2948;%%
 
\bibitem{sukhatme82}
U.P. Sukhatme and G. Wilk, Phys. Rev. D {\bf 25}, 1978 (1982).
%% CITATION = PHRVA,D25,1978;%%

\bibitem{lev83}
M. Lev and B. Petersson, Z. Phys. C {\bf 21}, 155 (1983).
%% CITATION = ZEPYA,C21,155;%% 
 
\bibitem{glauber59}
R.J. Glauber, in Lectures in Theoretical Physics, ed. W.E Brittin and L.G. Dunham, 
Interscience, N.Y., Vol. 1, p315 (1959); 
R.J. Glauber and G. Matthiae, Nucl. Phys. B {\bf 21}, 135 (1970).
%% CITATION = NUPHA,B21,135;%%

\bibitem{gribov}
V.N. Gribov, JETP {\bf 30}, 709 (1970) 
%% CITATION = SPHJA,30,709;%%

\bibitem{peitzmann99}
T. Peitzmann, Phys. Lett. B {\bf 450}, 7 (1999).
%% CITATION = PHLTA,B450,7;%%

\bibitem{gale99}
C. Gale, S. Jeon, and J. Kapusta, Phys. Rev. Lett. {\bf 82}, 1636 (1999).
%% CITATION = PRLTA,82,1636;%%

\bibitem{hufner00}
J. H{\"u}fner, B. Kopeliovich, and A. Polleri, hep-ph/0012010, hep-ph/0010282;
A. Polleri, nucl-th/0002054 (2000).
%% CITATION = HEP-PH 0012010;%%
%% CITATION = HEP-PH 0010282;%%
%% CITATION = NUCL-TH 0002054;%%

\bibitem{wang91}
X.N. Wang and M. Gyulassy, Phys. Rev. D {\bf 44}, 3501 (1991).
%% CITATION = PHRVA,D44,3501;%%

\bibitem{eskola99}
K.J. Eskola, V.J. Kolhinen, and C.A. Salgado, Eur. Phys. J. C {\bf 9}, 61 (1999).
%% CITATION = EPHJA,C9,61;%%

\bibitem{doksh01}
Yu.L. Dokshitzer, Phil. Trans. R. Soc. Lond. A {\bf 359}, 309 (2001).
%% CITATION = HEP-PH 0106348;%%

\bibitem{WA80} 
R. Albrecht {\it et al.} (WA80), Phys. Lett. B {\bf 361}, 14 (1995).
%% CITATION = PHLTA,B361,14;%%

\bibitem{WA98}
M.M. Aggarwal {\it et al.} (WA98), Phys. Rev. Lett. {\bf 85}, 3595 (2000); 
T. Peitzmann, private communication; nucl-ex/0006007; nucl-ex/0108006.
%% CITATION = PRLTA,85,3595;%%
%% CITATION = NUCL-EX 0006007;%%
%% CITATION = NUCL-EX 0108006;%%

\bibitem{david00}
G. David for the PHENIX collaboration, in
Proceedings of {\it Quark Matter 2001}, Nucl. Phys. A {\bf xxx}, 1 (2001);
http://www.rhic.bnl.gov/qm2001

\bibitem{levai01}
P. Levai, G. Papp, G. Fai, M. Gyulassy, nucl-th/0012017;
P. Levai, G. Papp, G. Fai, M. Gyulassy, G.G. Barnaf\"oldi, I, Vitev, 
and Y. Zhang, in Proceedings of {\it Quark Matter 2001}, 
Nucl. Phys. A {\bf xxx}, 1 (2001) (nucl-th/0104035);
http://www.rhic.bnl.gov/qm2001
%% CITATION = NUCL-TH 0012017;%%
%% CITATION = NUCL-TH 0104035;%%

\bibitem{BKKII}
J. Binnewies, B.A. Kniehl, and G. Kramer, Phys. Rev. D {\bf 52}, 4947 (1995).
%% CITATION = PHRVA,D52,4947;%%

\end{thebibliography}
\end{document}